# Effect of Size Asymmetry and Zwitterionic Character of Solvent Ions on Impedance Spectra of Ultradilute Colloidal Suspensions

Meenakshi[1], Shalini Gupta[2] and Gaurav Goel[1*]

[1] Dept. of Chemical Engineering, Indian Institute of Technology Delhi, 110016, India

[2]Asima Health Inc., 151 Charles St. West, Suite #199, Kitchener, ON N2G 1H6, Canada

**Corresponding author:* goelg@chemical.iitd.ac.in

**Abstract.** Buffer chemistry modulates colloidal impedance critically for label-free biosensing and electrokinetic diagnostics, yet classical models neglect buffer-specific interactions. We studied carboxylated polystyrene (PS) microspheres in conductivity-matched zwitterionic 4-(2-hydroxyethyl)-1-piperazineethanesulfonic acid (HEPES) and phosphate-buffered saline (PBS). Two concentration-dependent regimes emerged: oscillatory at ultradilute concentrations (<0.02 w/v%; Region 1) and monotonic at higher concentrations (Region 2). In Region 1, HEPES at pH 6.9 (approximately 80% zwitterionic) produced the largest deviation of delta Z from the O'Brien prediction. The conductivity increment was opposite in sign to classical expectation. We attribute this to zwitterion-driven interfacial ordering that enhances local conductivity and promotes self-limiting particle clustering. Dynamic light scattering detected cluster formation at 0.008 w/v%, well below aggregation thresholds. Changing pH from 6.9 to 8.1 strongly altered impedance amplitude; DLS profiles remained invariant. This indicates zwitterions modulate electrochemical polarization, not physical assembly. Molecular dynamics simulations showed pH-dependent HEPES organization at the PS surface. At pH 6.9, interfacial charge- and number-density profiles showed comparable oscillatory periodicity. At pH 8.1, anionic HEPES produced a more diffuse profile with increased Na+ compensation. These results support interfacial ordering as a contributor to enhanced conductivity but do not establish discrete layers or registry. Zeta potentials distinguished stable HEPES interfaces from PBS, where magnitude increased with ionic strength, contrary to classical screening theory. These nonlinear signatures and mesoscale organization underscore the role of buffer selection in high-sensitivity impedance biosensing.

## 1. Introduction

Zwitterionic molecules have spatially separated positive and negative charges and display strong hydrophilicity [1] and biocompatibility [2], making them promising candidates for various electrochemical applications. The fixed internal charge separation in these molecules leads to unique dielectric and solvation behaviors that significantly affect double-layer capacitance [3], interfacial interactions [4], and frequency-dependent responses in electrochemical systems [5]. At the electrode–electrolyte interface, zwitterions stabilize the electric double layer (EDL) [6, 7], reduce non-specific adsorption [6], and enhance overall conductivity [8]. These features directly contribute to increased reliability and sensitivity in electrochemical sensing techniques, especially those utilizing electrochemical impedance spectroscopy (EIS).

While solutions with zwitterionic species exhibit low intrinsic conductivity due to their overall neutrality, recent advancements in molecular design—such as enhanced charge delocalization [9] and engineered ion migration pathways [10]—have led to significant increases in ionic mobility. These strategies allow for precise control over the EDL structure and efficient modulation of interfacial charge transfer [11, 12]. When used as buffers or electrolytes in colloidal systems, the unique molecular architecture [14] and inherent ion-size asymmetry of zwitterions enhance dispersion stability [13] and yield distinctive impedance signatures across varying electrochemical conditions [11, 14].

EIS is a well-established technique for analyzing the behavior of colloidal suspensions, largely due to its sensitivity to particle surface properties, specifically surface potential [15]. At low frequencies, EIS measurements capture long-range interactions, dynamic aggregation, and characteristics of the diffuse EDL [17, 18]. In contrast, higher frequencies reveal the rapid migration and polarization of ions within the compact Stern and Helmholtz layers [16, 19]. This multifrequency approach elucidates the effect of particle charge, concentration, ion size, and electrolyte composition on the dielectric and conductive properties of colloidal systems [21, 22], allowing for direct observation of nanoscale structural or dynamic transformations driven by environmental changes [23].

Consequently, impedance measurements provide a detailed, frequency-resolved perspective on system properties while quantifying the macroscopic polarizability of colloidal suspensions based on the coupled response between the particle core and the surrounding ionic media [5]. Classical theories, grounded in Maxwell's analysis, predicted that increasing the volume fraction of insulating spheres within a conducting matrix would decrease overall system conductivity [24-27]. However, recent model extensions that consider solvent structure [28], ionic strength [29], and, importantly, ion-size asymmetry [29] have shown that system impedance and EDL capacitance in complex fluids are much more sensitive to molecular and ionic details [30, 31] than previously understood.

This increased sensitivity is particularly evident in zwitterionic and other non-symmetric buffer systems, where linear or mean-field approaches fall short. Accurate interpretation of EIS data requires an understanding of local microstructure and dynamic fluctuations [32]. Expanded models, drawing on observations from colloidal suspensions in ionic liquids at infinite dilution, demonstrate that ion-size asymmetry and local heterogeneities lead to significant deviations from classical expectations [33]. Larger ions induce uneven charge distributions and unique dipole accumulation within the EDL, altering interfacial electrostatic interactions [21]. Furthermore, localized ionic clusters form, intensifying these effects and significantly influencing ion mobility and interaction patterns at colloidal interfaces [21]. Consequently, electrokinetic behavior becomes highly dependent on the environment, with sensitive signal features reflecting the complex interplay between solvent architecture [34, 35], ionic asymmetry [36], and particle-interface dynamics [37, 38]—highlighting the need for a mechanistic framework based on molecular properties.

Within this broad context, colloidal systems and EDL theory are particularly important: the intricate relationship among particle charge, ionic atmosphere, and EDL structure predominantly influences the observed impedance behavior [39, 40]. Recent efforts to systematically extend classical theory investigate how ion-size asymmetry [41, 42] and overall charge neutrality [43, 44]—key characteristics of zwitterionic and ionic liquid systems—can lead to highly nonlinear and sensitive responses in EIS spectra. The dynamic interaction between highly mobile ions [45] and spatially heterogeneous environments [46], often found in both ionic liquids and certain zwitterionic solvents, offers a foundation for precise manipulation of signal response and sensitivity, even at ultra-low particle concentrations [47]. This integrative perspective emphasizes the vital role of particle-electrolyte interactions and microenvironmental conditions in the rational design and interpretation of EIS-based biosensing methodologies.

Several mechanistic questions remain unresolved. A key issue is understanding the relationship between solvent-ion configuration at a charged surface and the resulting impedance response in ultradilute suspensions. Unlike previous studies that focused on optimizing electrodes or increasing buffer ionic strength to manage double-layer effects [48-52], this work investigates the effect of the composition and physicochemical properties of the background medium. By systematically examining interfacial electrostatics, particle dynamics, and impedance responses in ultradilute polystyrene microsphere systems using classical electrolyte benchmarks, this study seeks to clarify how zwitterionic characteristics and ion-size asymmetry influence EIS sensitivity and selectivity.

## 2. Theory: Impedance Spectra of Colloidal Suspensions in Ionic Buffers

The impedance behaviour of charged particles in ionic solvents is strongly influenced by the structure of the electric double layer (EDL) and the polarization effects arising within the ionic cloud surrounding each particle [53, 54]. Building on foundational electrokinetic frameworks, Saville and O'Brien's model incorporated solution conductivity, ion polarization, electromigration, and electroosmosis, capturing both positive and negative changes in conductivity, depending on the particle's zeta potential and Debye length $\left(\frac{1}{\kappa}\right)$ [55, 56]. O'Brien's approach employed linear superposition for the limiting case of non-interacting particles and point-like ions [55], while Mangelsdorf and White extended the analysis to include frequency-dependent polarization, demonstrating that both free charges and displacement currents determine a system's complex conductivity under an applied AC field [26].

Recent advances in theory and experiments have improved the understanding of dielectric dispersion within concentrated colloidal suspensions, which play an increasingly important role in sensing and electrochemical applications [57]. The complex impedance $(Z_s(\omega))$ of these suspensions is commonly interpreted via dielectric and conductivity increments relative to the suspending medium, emphasizing the interplay between particle interactions, electric field effects, and solvent properties [37]. For systems with blocking electrodes, the complex conductivity of a suspension $\left(K_s^*(\omega)\right)$ is related to its experimentally measured impedance value $(Z_s(\omega))$, as per equation 1 [58].

$$\frac{1}{C_c Z_s(\omega)} = K_s^*(\omega) = i\omega\varepsilon_0 \varepsilon_s^*(\omega) = i\omega\varepsilon_0 \varepsilon_s^{'}(\omega) + i\omega\varepsilon_0 \varepsilon_s^{''}(\omega) \tag{1}$$

Here, $C_C$ is the cell constant, $\varepsilon_0$ is the vacuum permittivity, and $\varepsilon_s^*(\omega)$ is the complex relative permittivity of the suspension, comprising the real $\left(\varepsilon_s^{'}(\omega)\right)$ and imaginary $\left(\varepsilon_s^{''}(\omega)\right)$ components. The dielectric increment—quantifying the contribution of the suspended phase to the system's overall dielectric response—is:

$$\Delta\varepsilon_s^{'}(\omega) = \varepsilon_s^{'}(\omega) - \varepsilon_e^{'}$$

$$\Delta\varepsilon_s^{''}(\omega) = \varepsilon_s^{''}(\omega) - \varepsilon_e^{''} \tag{2}$$

where subscripts $e$ refers to the suspending medium. To incorporate zwitterionic effects, the mean-field field-theoretical model proposed by Bruch and Wang [59] represented each zwitterion molecule as an oligomer with oppositely charged terminal groups and $N$-1 neutral segments. The model considers system free-energy contributions from electrostatic interactions, chain entropy, and incompressibility. Under conditions relevant for AC impedance measurements (small, applied potentials), the effective dielectric constant was obtained as per equation 3:

$$\varepsilon_{eff} = \varepsilon_\infty + \frac{\beta e^2 \rho_0}{3\varepsilon_0} \cdot \frac{Nb^2}{N+1} \tag{3}$$

Here, $\varepsilon_\infty$ is the high-frequency (optical) dielectric constant from electronic polarization (the base contribution, excluding slower orientational dipole effects), $\rho_0$ is the zwitterion number density, $N$ is the chain length, $b$ is the segment length, $e$ is the elementary charge, and $\beta = \left(\frac{1}{k_B T}\right)$. This expression shows that $\varepsilon_{eff}$ increases with both zwitterion concentration and chain length, reflecting the enhanced dipole moment and polarizability induced by the zwitterionic chains. In colloidal suspensions, such enhancements are expected to modify ion organization and interfacial polarization, strongly impacting low-frequency spectra and the differential impedance signal ($\Delta Z$) [60].

Although experiments have confirmed that zwitterion addition increases dielectric constant [60, 61] and thus alters impedance [62], the precise details of how this molecular reorganization reshapes the surrounding ionic cloud remain an active area of research. Therefore, the observed reduction in impedance and improved measurement sensitivity can be interpreted as robust outcomes of zwitterion-driven electrostatic and dielectric phenomena, supported by both existing theory and experimental trends, while ongoing studies aim to unravel the underlying mechanisms fully.

## 3. Materials and Methods

**3.1** ***Materials.*** Platinum-on-glass planar interdigitated electrodes (IDEs) consisting of 40 symmetrical fingers (20 pairs), each 3.5 mm in length and 100 µm in width, with 100 µm edge-to-edge spacing, were used (Micrux Technologies, Spain). The electrodes were fabricated using high-purity platinum (≥99.99%, electronic grade) deposited on a glass substrate. Carboxylate-modified polystyrene (PS) beads, 1 µm in diameter with a surface charge density of 0.16 $Cm^{-2}$, were obtained from IDC Invitrogen (USA). Sodium hydroxide ($NaOH$, analytical grade, Merck, India) and acetone (≥99.5%, HPLC grade, Fisher Scientific) were used for pH adjustment and cleaning, respectively. The zwitterionic buffer 4-(2-hydroxyethyl)-1-piperazineethanesulfonic acid (HEPES, ≥99%, molecular biology grade, Sigma-Aldrich), with the molecular formula $C_8H_{18}N_2O_4S$ and $pK_a \approx 7.5$ at 25°C, was used to prepare buffer solutions. Phosphate-buffered saline (PBS) tablets (Omnipure) were composed of 13mM NaCl, 2mM KCl, and 10 mM phosphate buffer $(Na_2HPO_4/KH_2PO_4)$, pH 7.4 at 25°C. Polydimethylsiloxane (PDMS) base and curing agent (Sylgard 184, elastomer kit, Dow Corning, USA) were used for microfluidic sealing. Silicone grease (Metroark, laboratory grade) was applied to prevent fluid leakage. Ultrapure deionized (DI) water (~18.2 MΩ.cm resistivity at 25°C, Milli-Q system, Millipore, India) was used for all solution preparations. All reagents were used as received without further purification unless otherwise specified.

***3.2 Methods. (i) Sample Preparation.*** HEPES buffers (15, 40, and 80 mM) were prepared by dissolving HEPES free acid in ultrapure Milli-Q water (resistivity ~18.2 at 25°C). For example, 15 mM HEPES was prepared by dissolving 143 mg of HEPES in 40 mL of water, with the pH adjusted to 6.9, 7.4, or 8.1 using 0.1 M NaOH. 1× PBS stock was prepared by dissolving one PBS tablet in 100 mL of Milli-Q water and then diluted to match the conductivities of the respective HEPES buffers. All buffers were made fresh according to standard lab procedures, and all measurements were taken at room temperature (RT ~25°C). pH and conductivity were verified using a calibrated pH-7200 meter (Eutech) and a Lab-731 conductivity meter (Mettler Toledo). Carboxylate-modified polystyrene (PS) beads (IDC Invitrogen, USA; diameter 1; surface charge density 0.16) were prepared by centrifuging 1 mL of 4 w/v% stock at 10,375 g for 10 minutes. The pellets were washed three times with Milli-Q water to remove surfactants and preservatives before being resuspended in the target buffer and diluted to concentrations ranging from 0.002 to 0.1 w/v%.

***(ii) PDMS microchamber preparation.*** PDMS sheets (1.6 mm thick) were prepared using standard protocols [64] and cut into 1.5 cm x 1.2 cm pieces. A 4 mm diameter hole was punched through each piece to create the sample reservoir. Chambers were sealed with Scotch tape for storage. After each experiment, PDMS chambers were cleaned sequentially with 15 mM HEPES, 1 M NaOH, Milli-Q water, and acetone, then dried and resealed for reuse.

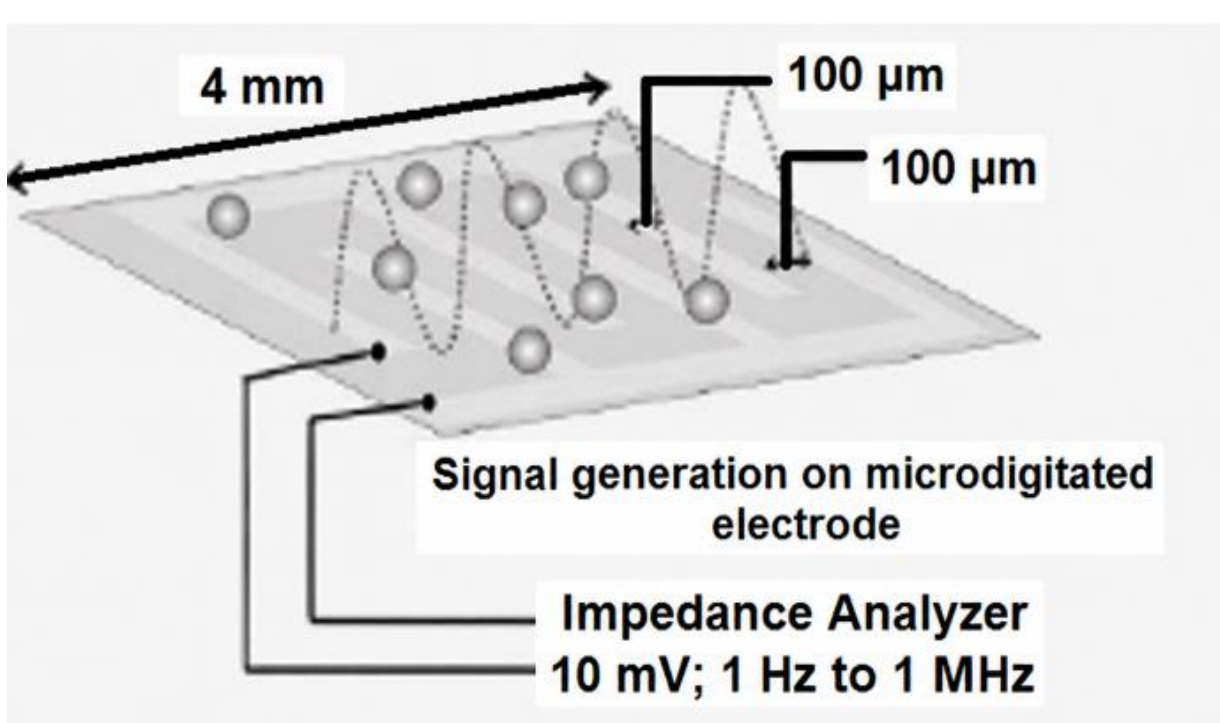

*Figure 1: Experimental setup consists of an interdigitated electrode (IDE) connected to an impedance analyzer to measure the impedance response, Z(ω) [37].*

***(iii) Experimental setup.*** The electrical measurements were performed using a high-precision impedance analyzer (Keysight E4990A) connected to planar platinum interdigitated electrodes (IDEs) on glass substrates [63]. The PDMS microchamber was mounted onto the IDEs, filled with the sample, and sealed with a glass coverslip using vacuum grease to prevent evaporation (Figure 1).

***(iv) Impedance measurements.*** For each experiment, 10 µL of the sample was loaded into the chamber, which was then covered with a glass coverslip. The sample was allowed to stabilize for 1 minute before recording the impedance at 10 mV and 1 MHz frequency using the impedance analyzer. Each suspension measurement was preceded by a buffer-only baseline measurement to minimize inter-assay variability. Since the phase angle remained consistently below 3° (indicating primarily resistive behavior), results are reported as differential impedance ($\Delta Z$) and signal-to-noise ratio (S/N). The differential impedance was calculated by subtracting the signal obtained from the buffer alone from the total signal measured for the suspension:

$$\Delta Z = Z_S - Z_{ref} \tag{4}$$

$$\frac{S}{N} = \frac{|\Delta Z|}{SD} \tag{5}$$

where $Z_S$ and $Z_{ref}$ represent the measured impedance of the colloidal suspension and the background buffer, respectively, and SD is the standard deviation of the measurement. Subsequently, normalizing these impedance changes to noise levels (S/N) provides a reliable benchmark for comparing signal clarity across buffer systems and distinguishing analyte-induced signals from background noise. All impedance measurements were conducted in triplicate $(n = 3)$ using independently prepared samples, and data are presented as mean ± SD. Following each test, the IDEs were thoroughly cleaned with buffer, acetone, and deionized (DI) water.

***(v) Sample characterisation.*** Particle morphology was determined using optical microscopy (Olympus BX53 with Orca Flash 4.0 Hamamatsu CCD camera) and scanning electron microscopy (SEM, Zeiss EVO 50). Surface and hydrodynamic properties were characterized in the respective buffers at 25°C. Zeta potential was measured using a Malvern Zetasizer MS-602 in folded capillary cells, with values calculated using the Smoluchowski model. Hydrodynamic diameter and size distribution were determined via Dynamic Light Scattering (DLS, Zeta Sizer NanoZS). Prior to analysis, all suspensions were sonicated and vortexed alternately for 2 minutes to ensure uniform dispersion. Measurements were performed in triplicate ($n = 3$), and data are reported as mean ± SD.

***(vi) Molecular dynamics simulations.*** All-atom simulations were performed using GROMACS 2020.3 [64] with the CHARMM36 force field [65] and TIP3P water [66] to investigate the HEPES-mediated electrical double layer at the negatively charged polystyrene interface. Simulations were performed for 15 mM HEPES at pH 6.9 and 8.1 in contact with a polystyrene slab of surface charge density −0.20 $e \cdot nm^{-2}$; detailed system construction, equilibration, production, and analysis procedures are provided in Supplementary Information, Section S.5.

## 4. Results

The impedance behaviour of carboxylated polystyrene colloidal (PS-latex) suspensions was systematically studied for various particle concentrations and buffer compositions in both HEPES and conductivity-matched PBS solutions.

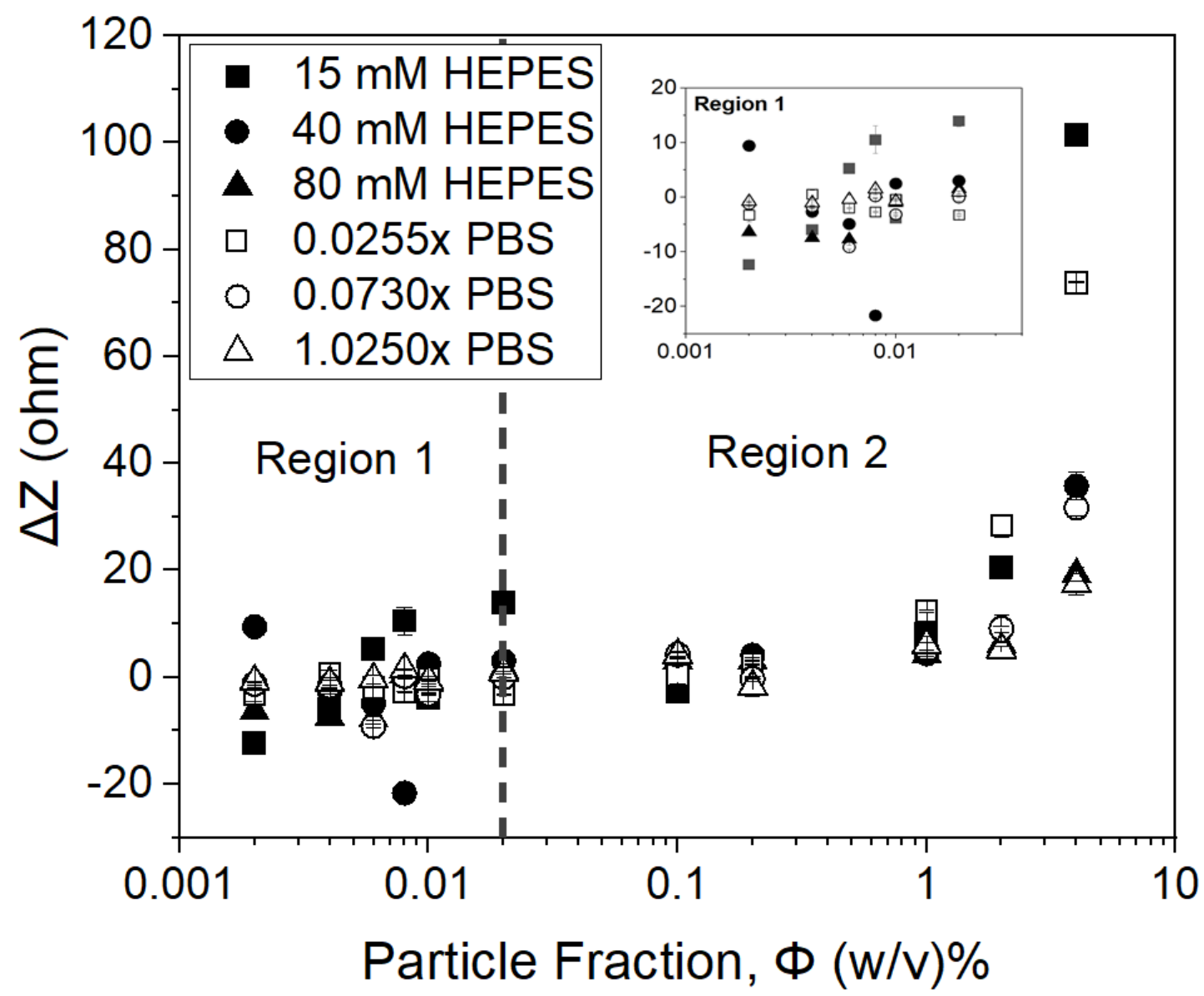

*Figure 2: The change in impedance (ΔZ) of PS-latex particle suspensions was measured at 1 MHz for different Φ (w/v%) in HEPES buffers (solid symbols: 15 mM, 40 mM, 80 mM) and conductivity-matched PBS buffers (hollow symbols: 0.0255x, 0.0730x, 0.1250x) at pH 7.4. The inset shows a magnified view of ΔZ in Region 1. SDs where not visible are smaller than the data points.*

Figure 2 shows the dependence of impedance change (ΔZ) on particle volume fraction (Φ) for PS-latex suspensions at pH 7.4. A comparison between HEPES and conductivity-matched PBS reveals two distinct dielectric regimes. In Region 1 (Φ < 0.02 w/v%), the HEPES suspensions exhibited an oscillatory ΔZ response, with a pronounced local maximum near the upper boundary of this region (approximately 0.02 w/v%), whereas the corresponding conductivity-matched PBS suspensions displayed a substantially weaker response. This anomalous sensitivity at ultra-low concentrations is most prominent in the 15 mM HEPES system. Notably, while the buffers are matched for bulk conductivity, the magnitude and profile of ΔZ in HEPES deviate significantly from the PBS controls. In Region 2 (Φ > 0.02 w/v%), the impedance response transitions to a near-monotonic increase with Φ. In this higher-concentration regime, the responses of HEPES and PBS converge, suggesting that as the particle density increases, the influence of bulk conductivity begins to dominate buffer-specific chemical effects. These observations indicate that the enhanced sensitivity in HEPES is driven by packing and organization of the zwitterionic species at particle surface rather than macroscopic electrolytic properties. These findings are consistent with Khandelwal et al. [37], who reported that zwitterionic HEPES buffers generated enhanced impedance amplification and sharper transitions compared to PBS under equivalent conductivity conditions.

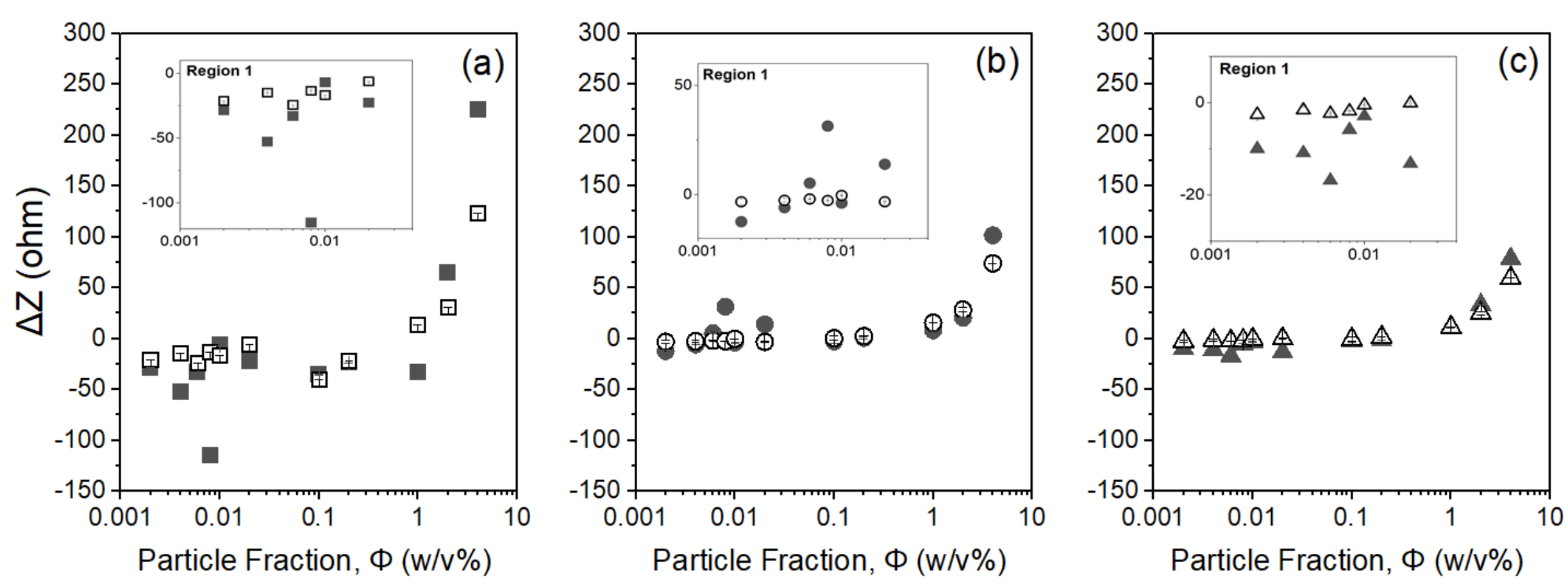

*Figure. 3: Impedance response (ΔZ) of latex particles suspended in 15 mM HEPES (solid) and its corresponding conductivity-matched PBS (hollow) buffer at 1 MHz and at (a) pH 6.9, (b) pH 7.4, and (c) pH 8.1, plotted against Φ (w/v%). Insets show an expanded view of the low particle fraction regime, highlighting the oscillatory features observed in the HEPES suspensions.*

To further isolate the role of the zwitterionic species, Figure 3 examines the influence of pH modulation on the impedance response. A significant suppression of the oscillatory features is observed as the pH increases from 6.9 to 7.4, resulting primarily from the reduced fraction of zwitterionic species, which are crucial for facilitating particle-specific interfacial interactions. At pH values exceeding 7.4, the increased ionic strength and bulk conductivity dominate the system's dielectric response, leading to the near-complete convergence of the HEPES and PBS impedance profiles. Although the lower zwitterion content at higher pH slightly impacts the overall S/N, its primary effect is the elimination of the distinctive oscillatory features in Region 1. This confirms that zwitterion-mediated effects are the primary drivers of the observed low-concentration anomalies. These results align with broader literature; for example, Shah et al. [14], Khandelwal et al. [37], and Hallett et al. [67] have demonstrated that high-sensitivity impedance responses in colloidal systems are uniquely coupled to the presence of zwitterionic species.

Table 1: Impedance response (ΔZ) and signal-to-noise ratios (S/N) for PS-latex particles measured in various HEPES and conductivity-matched PBS buffers at 1 MHz. Values represent the mean ± SD from three independent replicates at each particle concentration.

| **Buffer** | **Signal/Noise** | | $\frac{d^2(\Delta Z)}{d\Phi^2}$ **(Ω × 10$^{-7}$)** | |
|---|---|---|---|---|
| | **Region 1** | **Region 2*** | **Region 1** | **Region 2*** |
| 15 mM HEPES; pH 6.9 | 35.82 ± 0.45 | 39.24 ± 0.23 | 3.954 ± 0.008 | 0.226 ± 0.009 |
| 0.0139x PBS; pH 7.4 | 22.14 ± 0.38 | 24.54 ± 0.28 | 0.682 ± 0.002 | 0.005 ± 0.003 |
| 15 mM HEPES; pH 7.4 | 25.01 ± 0.34 | 28.66 ± 0.31 | 1.585 ± 0.007 | 0.072 ± 0.001 |
| 0.0255x PBS; pH 7.4 | 19.53 ± 0.26 | 21.77 ± 0.41 | 0.310 ± 0.001 | 0.004 ± 0.004 |
| 15 mM HEPES; pH 8.1 | 24.70 ± 0.66 | 26.61 ± 0.22 | 0.583 ± 0.002 | 0.006 ± 0.002 |
| 0.0490x PBS; pH 7.4 | 14.37 ± 0.41 | 19.62 ± 0.14 | 0.263 ± 0.004 | 0.002 ± 0.002 |
| 40 mM HEPES; pH 7.4 | 20.37 ± 0.53 | 22.28 ± 0.22 | 1.113 ± 0.001 | 0.050 ± 0.005 |
| 0.0730x PBS; pH 7.4 | 4.99 ± 0.19 | 12.19 ± 0.17 | 0.458 ± 0.003 | 0.003 ± 0.001 |
| 80 mM HEPES; pH 7.4 | 14.42 ± 0.23 | 18.98 ± 0.19 | 0.380 ± 0.002 | 0.004 ± 0.001 |
| 0.1250x PBS; pH 7.4 | 4.27 ± 0.08 | 17.81 ± 0.11 | 0.120 ± 0.001 | 0.003 ± 0.003 |

*Region 2 calculations are limited to a particle fraction (Φ) of 2 w/v%, as the final data point at 4 w/v% was identified as an outlier and excluded from the average.

The non-monotonic relationship between particle concentration (Φ) and impedance *ΔZ(Φ)* was assessed using the slope and curvature of the response curve, while relevance to sensing applications was quantified through the signal-to-noise ratio (S/N) (Table 1). The colloidal suspension in 15 mM HEPES buffer at pH 6.9 exhibited the highest S/N ratios in both regions, with pronounced oscillations with Φ captured by the highest $\frac{d^2(\Delta Z)}{d\Phi^2}$ (curvature) in region 1. In contrast, PBS-based suspensions showed consistently lower and less distinct values. These results align with findings from Kelly et al. [68], who noted that PBS has lower S/N ratios and higher electrical noise due to its greater ionic mobility and conductivity, in contrast to the enhanced signal clarity and colloid stability provided by zwitterionic buffers like HEPES. Similarly, Khandelwal et al. [37] found that the ion size asymmetry in HEPES enhances counterionic structuring, a stabilizing effect that is absent in PBS under similar conditions.

To investigate the molecular basis of the oscillatory ΔZ response, atomistic molecular dynamics (MD) simulations were performed for the HEPES–polystyrene interface at pH 6.9 and 8.1. At pH 6.9, the zwitterionic form (HEPES(Z)) constituted approximately 80% of the HEPES population, with the anionic form (HEPES(A)) accounting for the remaining 20%; these proportions were approximately reversed at pH 8.1 (Figure S2). Figures 4 and 5 show the corresponding charge-density, $\rho_q(z)$, and normalized number- density profiles, $\bar{n}(z)$, for the zwitterionic, anionic, and $Na^+$ components, plotted as a function of distance ($z$) from the negatively charged polystyrene surface. Together, these profiles characterize the pH-dependent interfacial organization of HEPES and provide the structural basis for interpreting the oscillatory ($\Delta Z$) response.

At pH 6.9, the HEPES(Z) charge-density profile shows alternating positive and negative features extending to approximately z ≈ 18 nm. Similar oscillations are observed in the normalized number-density profile, with an inter-peak spacing of approximately δ ≈ 0.7–0.9 nm, comparable to the characteristic molecular dimension of HEPES. This correspondence in periodicity between the two profiles suggests correlated molecular-scale organization near the polystyrene surface. At pH 8.1, the near-surface HEPES(Z) feature is retained but decays to within the noise level beyond z ≈ 6 nm, and the long-range oscillatory structure observed at pH 6.9 is no longer resolved. Conversely, the HEPES(A) contribution becomes more pronounced, reaching approximately –0.010 to –0.013 $q/nm^3$ at z ≈ 17–18 nm, while the $Na^+$ contribution increases to approximately +0.029–0.030 q/nm³. These changes indicate that the long-range oscillatory response is primarily associated with the larger zwitterionic HEPES population at pH 6.9. Because only total charge- and number-density profiles are considered here, the observed periodicity is interpreted as evidence of molecular-scale interfacial ordering rather than as proof of discrete single-molecular layers.

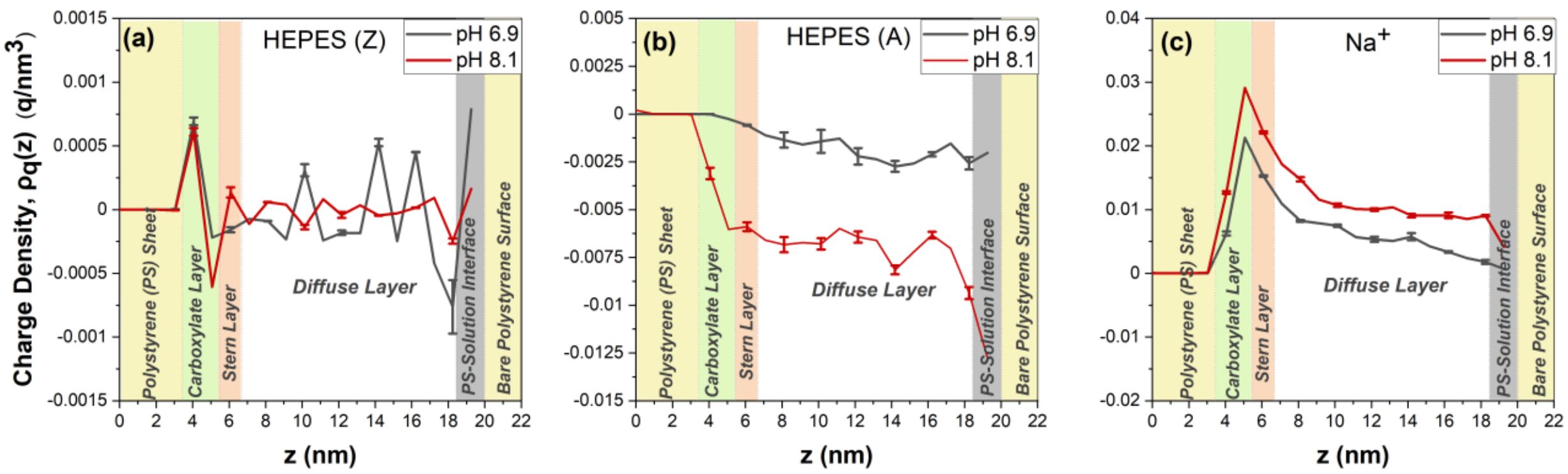

*Figure 4. Charge density profiles, $\rho_q(z)$, as a function of distance z from the polystyrene surface (σ = −0.20 e·nm⁻²): (a) HEPES zwitterion (Z), (b) HEPES anion (A), and (c) Na⁺ counterions at pH 6.9 (black) and pH 8.1 (red). Results are derived from atomistic MD simulations; error bars represent standard deviations from block averaging.*

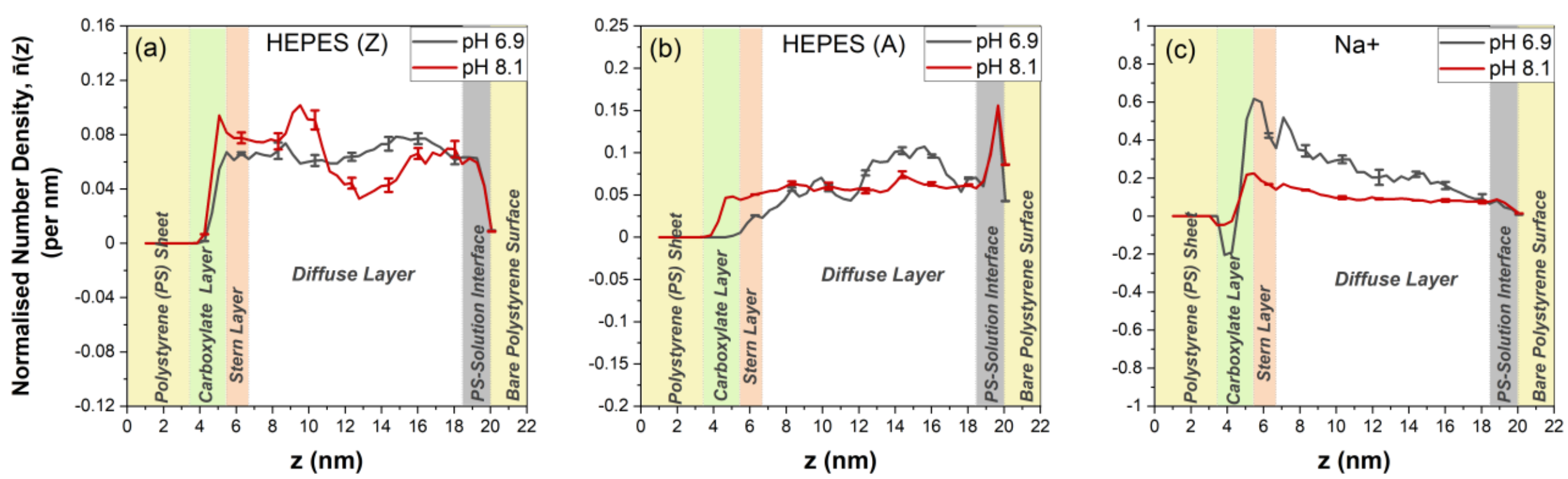

*Figure 5. Normalized number density profiles, $\bar{n}(z)$, as a function of distance z from the polystyrene surface (σ = −0.20 e·nm⁻²): (a) HEPES zwitterion (Z), (b) HEPES anion (A), and (c) Na⁺ counterions at pH 6.9 (black) and pH 8.1 (red). Results are derived from atomistic MD simulations; error bars represent standard deviations from block averaging.*

Table 2 summarizes the zeta potential (ξ) values for PS-latex particles suspended in HEPES and conductivity-matched PBS buffers across a range of pH levels. All measurements were performed at a low particle volume fraction (0.008 w/v%) to eliminate interparticle interference. Since the experimental pH range significantly exceeds the $pKa$ of the surface carboxyl groups, the particles remained fully ionized, ensuring that variations in $\xi$ were driven by buffer-ion interactions rather than changes in surface dissociation [69].

Table 2: HEPES buffers (15 mM at pH 6.9, 7.4, and 8.1) and corresponding conductivity-matched PBS buffers used in this study for 0.008 w/v% PS-latex suspensions, with measured zeta potentials.

| Buffer | Conductivity (S/m) | Zeta Potential (mV) |
|---|---|---|
| 15 mM HEPES; pH 6.9 | 0.026 | -63.38 ± 0.77 |
| 0.0139x PBS; pH 7.4 | | -12.91 ± 3.33 |
| 15 mM HEPES; pH 7.4 | 0.050 | -54.28 ± 5.68 |
| 0.0255x PBS; pH 7.4 | | -34.65 ± 3.78 |
| 15 mM HEPES; pH 8.1 | 0.095 | -49.6 ± 1.73 |
| 0.0490x PBS; pH 7.4 | | -48.2 ± 2.82 |

In HEPES buffers, the particles exhibited a consistently high negative zeta potential across all tested pH values. Notably, only a marginal decrease in magnitude was observed as pH (and consequently conductivity) increased. This stability highlights the robust ampholytic behaviour of HEPES [70]. In contrast, the PBS-based suspensions exhibited non-classical electrostatic behaviour. At lower ionic strengths, the zeta potential magnitude was significantly reduced compared to the HEPES environment. However, as the PBS concentration increased to match the highest conductivity point, the zeta potential magnitude surprisingly increased, eventually converging with the HEPES value. This observed increase in zeta potential with PBS concentration contradicts classical DLVO theory and the Helmholtz-Smoluchowski expectation of electrostatic screening [71]. This anomaly is supported by the work of Jorgensen et al. [72] and Garg et al. [73], who demonstrated that at very low ionic concentrations, specific ion-particle interactions can enhance effective surface charge through preferential adsorption or by shifting the position of the slipping plane.

To elucidate differential interfacial responses in the presence of zwitterions, we compared experimental impedance data with predictions from the O'Brien analytical model [56] (see Figure 6). This classical model predicts a monotonic increase in $\Delta Z$ with increasing buffer conductivity, based on the linear superposition of particle effects typical of simple electrolytes [74]. While both buffer systems exhibit deviations from this classical framework at ultradilute concentrations (<0.02 w/v%), the magnitude and nature of these discrepancies are significantly more pronounced in HEPES at pH 6.9. In this ultradilute regime, HEPES produces a distinctive, non-monotonic ΔZ response featuring a minimum that the O'Brien model fails to capture. Although the calculation incorporates essential parameters—including measured zeta potential (approximately -60 mV for HEPES and -12 mV for PBS), ionic mobility, and diffusion coefficients—these factors alone are insufficient to reconcile the theory with experimental observations. Consequently, the conductivity increment in HEPES-based suspensions (Figure 6) at these low concentrations exhibits an opposite sign and a substantially higher magnitude than classical expectations.

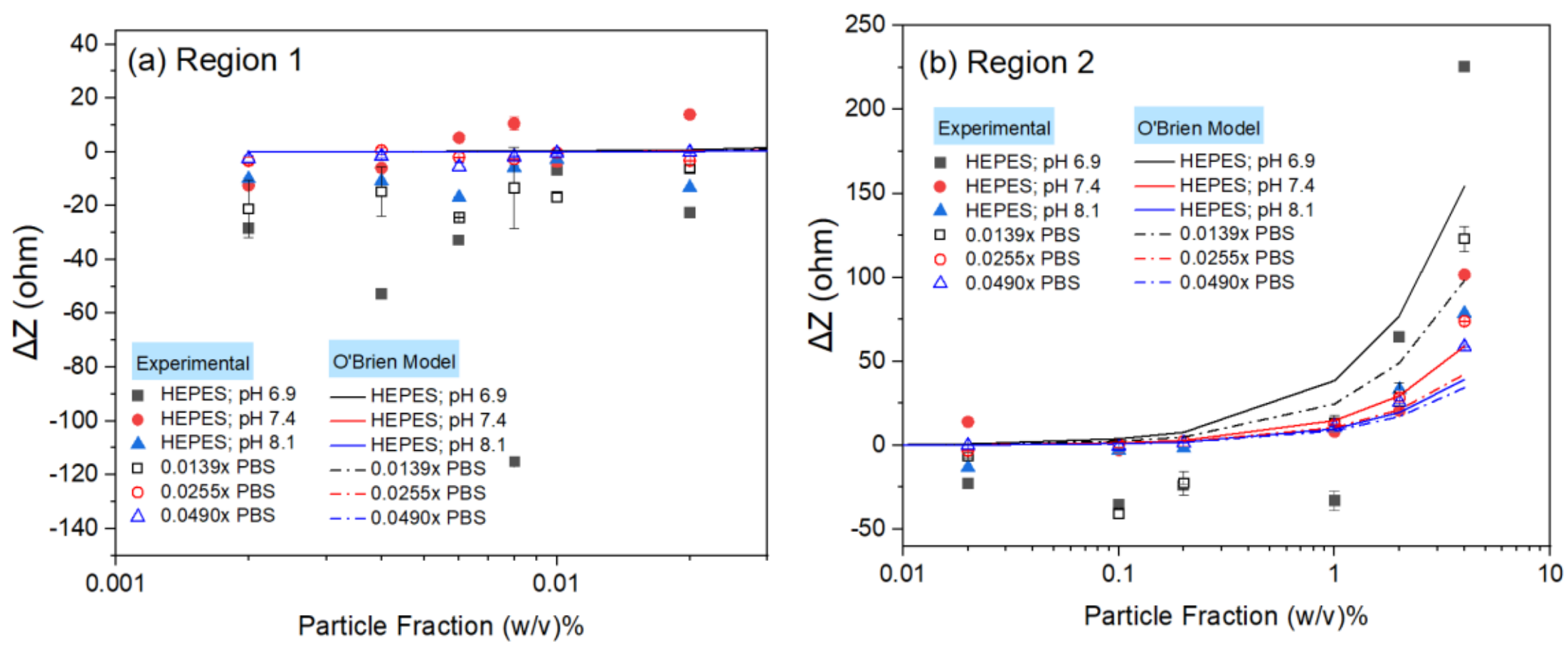

*Figure 6: Impedance change (ΔZ) as a function of PS-latex particle volume fraction (Φ) in 15 mM HEPES buffers (fixed concentration across pH conditions) and conductivity-matched PBS buffers (concentrations adjusted at pH 7.4) in (a) Region 1 and (b) Region 2. Solid symbols represent experimental data for HEPES at pH 6.9 (black), pH 7.4 (red), and pH 8.1 (blue); hollow symbols show PBS at matched conductivity 0.0139x (black), 0.0255x (red), and 0.0490x (blue). Solid lines represent O'Brien model predictions for 15 mM HEPES at pH 6.9 (black), pH 7.4 (red), and pH 8.1 (blue); dash-dotted lines correspond to PBS model predictions.*

Dynamic light scattering (DLS) analysis provides further evidence for complex interfacial dynamics. As ionic strength increases (Figure 7), a distinct shoulder emerges in the hydrodynamic diameter distribution. This shoulder becomes significantly more pronounced at 0.008 w/v%—nearly half the concentration where the impedance behaviour shifts to a monotonic increment—suggesting emergence of particle clusters. Interestingly, despite the significant differences in impedance between pH 6.9 and pH 8.1 (Figure 3), the DLS distributions (Figure 8) remain virtually identical across this pH range. This lack of change in the mean hydrodynamic diameter indicates that while the zwitterion fraction modulates the electrochemical response and conductivity at the interface, it does not necessarily result in permanent, large-scale structural changes to the clusters detectable by DLS in the ultradilute regime. Microscopy analysis (Figure S7, S8) confirms that clustering is not the dominant factor at the lowest concentrations, although subtle stable clusters begin to form as early as 0.008 w/v%.

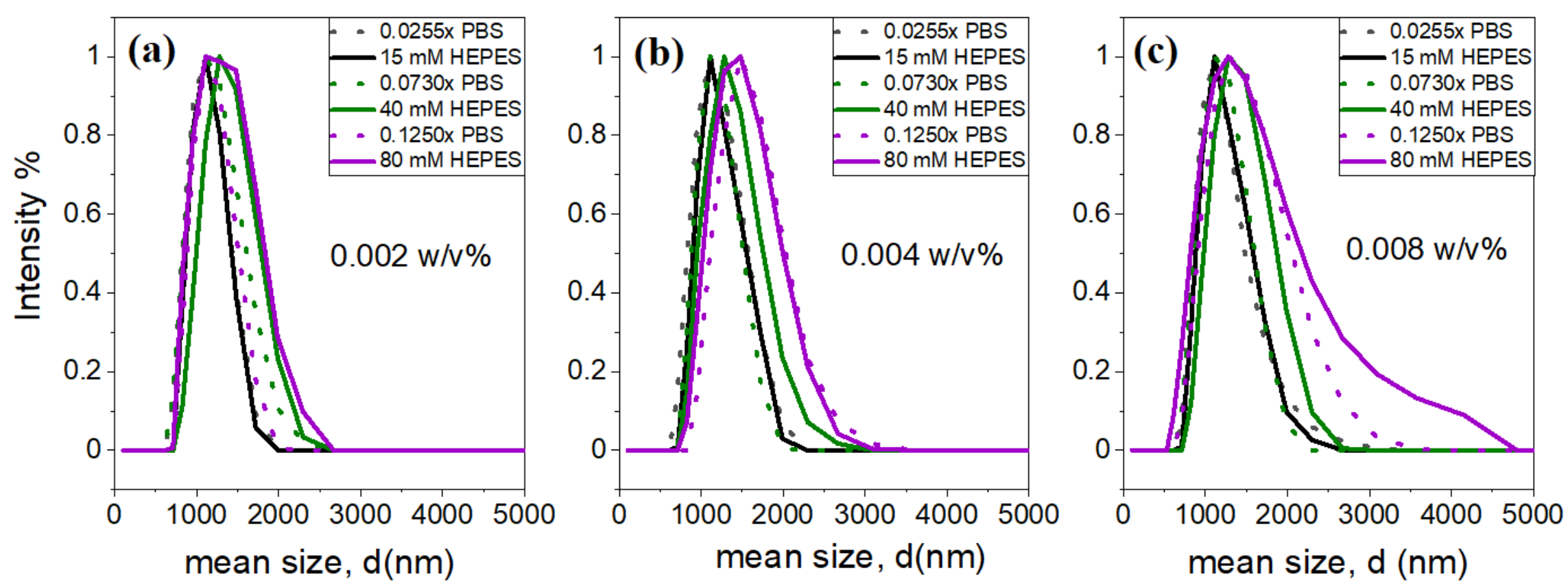

*Figure 7: Intensity-weighted particle size distributions for latex particle suspension in 15 mM (black), 40 mM (red), and 80 mM (blue) HEPES buffer (solid) at pH 7.4 and their conductivity-matched 0.0255x (black), 0.0730x (red), and 0.1250x (blue) PBS buffer (dotted) for three different particle fractions (a) 0.002 w/v%, (b) 0.004 w/v%, and (c) 0.008 w/v%. Profiles were obtained via DLS from homogenized suspensions. Data represent mean ± SD.*

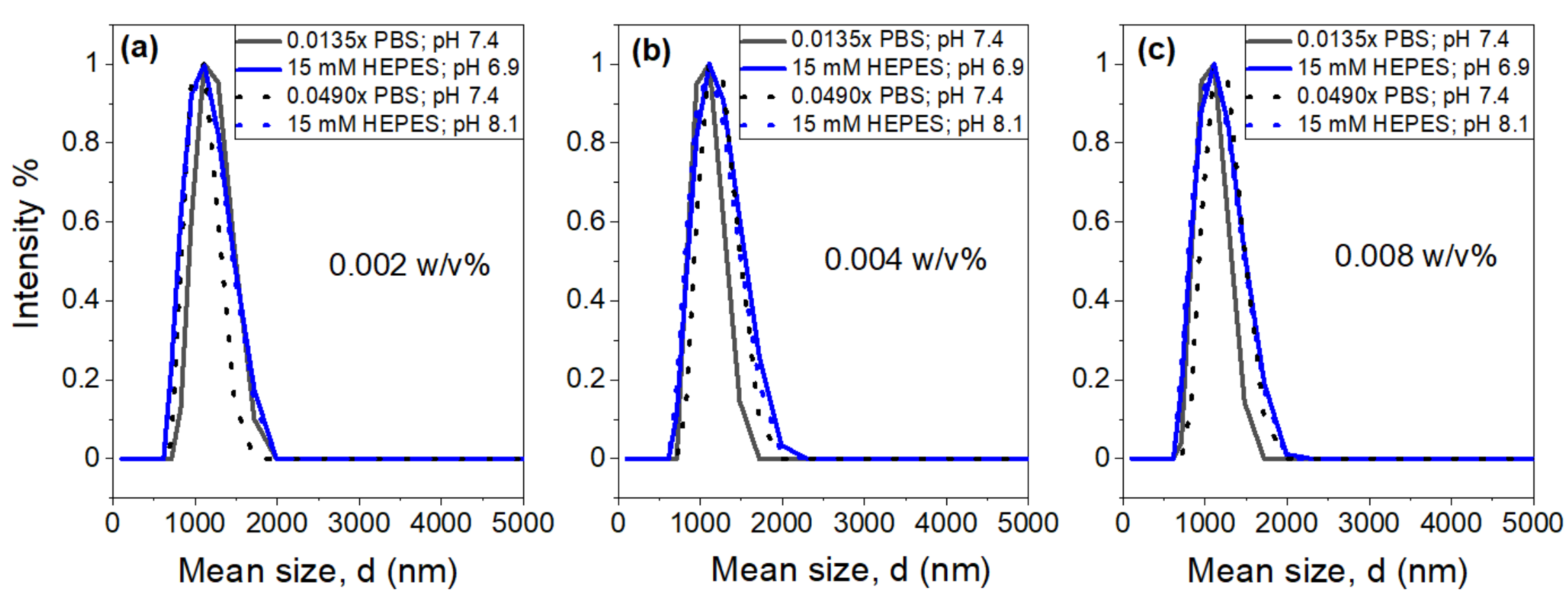

*Figure 8: Intensity-weighted particle size distributions of latex particle suspensions in 15 mM HEPES buffer at pH 6.9 (solid red line) and pH 8.1 (dashed red line), compared with conductivity-matched PBS buffers: 0.0135x PBS (solid black line) and 0.0490x PBS (dashed black line), for*

*three different particle fractions: (a) 0.002 w/v%, (b) 0.004 w/v%, and (c) 0.008 w/v%. Profiles were obtained via DLS from homogenized suspensions. Data represent mean ± SD.*

## 5. Discussion

The impedance behaviour of PS-latex particle suspensions in HEPES and conductivity-matched PBS exhibits two distinct regimes across varying particle concentrations: an oscillatory, nonlinear response at ultra-dilute concentrations ($\Phi < 0.02$ w/v%; Region 1) and a monotonic increase at higher particle loadings (Region 2). Figures 2 and 3 illustrate that the divergence between zwitterionic and electrolytic environments is most pronounced in Region 1, where $\Delta Z$ and its derivatives show significant differences (Table 1). In contrast, the responses converge in Region 2, suggesting a fundamental shift in the dominant electrokinetic interaction mechanisms as macroscopic electrolytic properties begin to dominate buffer-specific chemistry at higher particle densities [75].

The failure of classical electrokinetic frameworks, such as the O'Brien model [56], to capture the non-monotonic impedance behaviour in HEPES (Figure 6) underscores the limitations of standard screening theories in zwitterionic media. Specifically, the system exhibits an anomalous conductivity increment at ultra-low particle concentrations and low buffer conductivities—a highly nonlinear response that independent particle-based superposition models fail to predict. This effect is most pronounced in zwitterionic buffers at the lowest pH (pH 6.9)—where the zwitterionic fraction is maximal—and is significantly attenuated as pH increases or when transitioning to PBS controls. This discrepancy indicates that the enhanced $\Delta Z$ response is driven by intrinsic zwitterionic effects that transcend macroscopic electrolytic properties [4, 67]. As the ionic concentration increases—moving from HEPES to the higher ionic strength of PBS—the electrokinetic screening effect becomes the primary driver of the dielectric response [57]. The compression of the electrical double layer (EDL) effectively shields the particle's surface charge and reduces the characteristic length scale of ion migration [76]. This suppression of interparticle coupling leads to a collapse of the volume-fraction dependence of the dielectric increment—a phenomenon well-captured by standard models [77, 79] in Region 2, where the response scales linearly with particle number.

While the observed impedance modulation can be interpreted through extended electrokinetic frameworks—such as the Mangelsdorf and O'Brien models [77, 78] incorporating dynamic Stern layer (DSL) effects and stagnant-layer conductivity [80]—these remain insufficient to reconcile theory with zwitterionic experimental data. We propose that the superior signal clarity and oscillatory deviations observed in 15 mM HEPES at pH 6.9 (Table 1) arise from two synergistic mechanisms: (a) classical electrostatic screening, where the large Debye length ($\kappa^{-1}$) facilitates interparticle polarization, and (b) the formation of a highly structured interfacial layer around particles and small clusters via dipolar coordination [37, 81, 82]. Unlike the simple monovalent ions in PBS, the bulky, piperazine-based structure of HEPES promotes a responsive Stern layer and steric-like stabilization at the shear plane [83]. This zwitterion-facilitated hydrated layer prevents compact electrical double layer (EDL) compression [84], disrupts conventional screening, and enhances local ion mobility. Consequently, lateral ion transport and stagnant-layer conductivity [80] provide additional relaxation modes that intensify impedance sensitivity, particularly at lower frequencies [85].

This interfacial structuring is manifested in the primary impedance peak observed at 0.02 w/v% (Figure 2), which likely arises from electrokinetic overlap among isolated HEPES molecules or small zwitterionic assemblies (~1 nm scale) behaving as discrete electrokinetic entities. The impedance response is governed by the interplay between particle and zwitterion concentrations, revealing two distinct sub-regions within the oscillatory Region 1: (i) the infinitely dilute limit (< 0.003 w/v%), where independent particles exhibit enhanced ion mobility due to localized zwitterionic layering, and (ii) a transition zone (approximately 0.003 to 0.02 w/v%), where 2–3 particle clusters create optimal spatial arrangements that facilitate the development of extended zwitterionic networks [33, 86, 87]. These networks facilitate a conductivity increase that exceeds classical double-layer predictions [88] and cannot be reconciled with models assuming idealized, non-interacting spheres. We conjecture that the initial zwitterion-driven layering at the particle interface serves as the primary driver for the sensitive impedance response in Region 1, which subsequently nucleates the particle clustering observed at higher loadings. This process is inherently self-limiting: the same clustering, once it reaches a sufficient packing density, disrupts the zwitterionic order that initiated it, thereby driving the transition to Region 2 behaviour.

The oscillatory $\Delta Z$ response is attributed to the protonation state of HEPES. The pH-dependent deprotonation of the piperazine nitrogen ($pKa_2 \approx 7.5$) modulates the zwitterionic population [89], which in turn dictates the impedance signature. At pH 6.9, where HEPES(Z) is dominant, the similar periodicity in the charge-density and normalized number-density profiles indicates that the response reflects correlated molecular-scale structuring rather than a purely monotonic electrostatic diffuse layer [89–91]. The repeat distance of 0.7–0.9 nm is comparable to the molecular dimension of HEPES, providing a plausible structural basis for the oscillatory impedance response. At pH 8.1, the reduced HEPES(Z) population and increased contributions from HEPES(A) and $Na^+$ coincide with the disappearance of persistent HEPES oscillations beyond the near-surface region. This loss of molecular-scale ordering provides a structural explanation for the attenuation of $\Delta Z$ toward the PBS-control response (Table 1).

The oscillatory dependence of $\Delta Z$ on particle volume fraction, $\Phi$, is consistent with, but does not establish, a registry-based interpretation of the interfacial structure. As $\Phi$ changes, the mean interparticle separation, $d$, varies relative to the HEPES repeat distance, $\delta$. When $d$ approaches an integer multiple of $\delta$, interfacial density modulations at opposing surfaces could become more favourably registered, enhancing their contribution to the measured $\Delta Z$. At intermediate separations, geometrical frustration would reduce such registry and attenuate the response [92]. Because the present simulations contain only a single interface, phase matching between opposing surfaces cannot be evaluated directly. This mechanism should therefore be regarded as a possible interpretation of the experimental $\Delta Z$ trends rather than as a demonstrated molecular process. The subsequent evolution from porous, loosely connected aggregates [93, 94] to denser assemblies [95, 96] would reduce the available interfacial space and increase inter-cluster contacts. These changes are consistent with progressive perturbation of the confined interfacial HEPES structure and suppression of the oscillatory response [97], potentially contributing to the transition to Region 2 [98].

Mesoscale structuring evidenced by DLS corroborates this interpretation: the evolution of the hydrodynamic diameter distribution near 0.02 w/v% suggests that zwitterion-mediated assembly precedes large-scale aggregation and signals the onset of interparticle associations that dominate the high-concentration regime (Figures 7, 8) [99]. Crucially, the decoupling of impedance sensitivity from hydrodynamic size changes indicates that the zwitterion fraction primarily modulates electrochemical polarization and interfacial conductivity, rather than inducing permanent structural reorganization in the ultradilute regime [100]. These observations are supported by a growing body of literature emphasizing the unique role of zwitterionic environments. Khandelwal et al. [37] demonstrated that such collective, non-classical responses are characteristic of zwitterionic systems due to ion-size asymmetry and specific polarization effects. Similarly, Garg et al. [73] and Parsons et al. [101] highlight that interpreting interfacial data requires moving beyond classical screening theory to consider the specific electrostatic influences of buffer composition. The enhanced sensitivity observed here mirrors findings by Osarinmwian et al. [102], who noted that zwitterions could promote particle associations and modulate solution permittivity in ways standard electrolytes cannot. Together, these results provide a direct link between zwitterion-mediated layering and the resulting impedance anomalies, confirming that the spike in conductivity at low concentrations is a hallmark of zwitterion-driven interfacial phenomena that precede and eventually drive progressive particle clustering [103].

This complex behaviour points toward an optimal zwitterionic concentration for conductivity enhancement, which in this study occurs at 40 mM HEPES (Table 1). At lower concentrations (e.g., 15 mM), the particle surface may be insufficiently covered [104]. However, decreasing the pH from 7.4 to 6.9 at 15 mM HEPES increases the zwitterionic fraction, leading to enhanced oscillations and a greater conductivity increment, indicating larger surface interactions (Table 1, Figure S4).

Zeta potential measurements (Table 2) provide further insight into the distinct stabilization mechanisms in HEPES and PBS. The resilience of the HEPES zeta potential across varying pH and concentrations—characterized by a consistently high negative magnitude—suggests that stabilization is governed by dipolar interactions with zwitterionic species rather than classical ion-mediated screening [14, 67, 105]. This stability stems from the robust ionization profile of piperazine-based zwitterions, which preserves a pH-responsive yet stable electrostatic environment that resists the double-layer compression typical of simple salts [84]. In contrast, the PBS system remains highly sensitive to specific ion-surface affinities and ion-deficient instability at low concentrations, reflecting a transition from competitive adsorption to standard electrolyte screening [101]. Thus, while the zwitterionic resilience and steric bulk of HEPES maintain a stable interfacial potential, PBS behavior is dominated by buffer-specific effects that override simple valency-based models [83, 106].

These stabilization differences underpin the broader mechanistic transitions observed across the concentration spectrum. In the ultradilute regime (Region 1), the convergence of many-body interactions and zwitterion-mediated structuring creates nonlinearities that classical linear models fail to capture. As the system transitions to Region 2, the impedance response in HEPES is further intensified by aggregation that reduces the effective dielectric volume [107, 108], a phenomenon that contrasts with the simple additive scaling observed in PBS. Together, these observations confirm that the zwitterionic environment fundamentally alters the coupling between particle clusters and the surrounding electrolyte, shifting the dominant electrokinetic mechanism from interface-driven conductivity enhancement to clustering-dominated dielectric modulation.

## 6. Conclusion

This work demonstrates that the chemical nature of the buffering environment significantly influences colloidal impedance, with zwitterionic HEPES eliciting unique, nonlinear dielectric signatures that classical electrokinetic models fail to predict. By identifying two distinct concentration regimes, we show a mechanistic shift from interface-driven conductivity enhancement in the ultradilute Region 1 ($\Phi < 0.02$ w/v%) to clustering-dominated responses in Region 2. Crucially, the decoupling of impedance sensitivity from DLS hydrodynamic profiles confirms that zwitterions modulate electrochemical interfacial properties—likely via dynamic Stern layer effects—more significantly than they alter physical cluster structures. The pronounced, pH-dependent impedance response of HEPES aligns with its molecular organization at the polystyrene interface. At pH 6.9, where HEPES(Z) predominates, simulations reveal persistent molecular-scale interfacial oscillations. In contrast, at pH 8.1, the increased contributions from HEPES(A) and $Na^+$ coincide with an attenuation of these oscillations. Thus, the simulations provide molecular-level support for associating the enhanced HEPES impedance response at pH 6.9 with zwitterion-mediated interfacial ordering, though they do not establish a direct causal mechanism. These findings expose a critical gap in linear electrokinetic theory, which assumes idealized, non-interacting spheres and ignores buffer-specific molecular coordination. Our results necessitate the development of new theoretical frameworks that account for zwitterion-mediated layering and mesoscale organization. Such advancements are essential for accurately predicting and optimizing high-sensitivity, impedance-based colloidal sensing platforms at clinically relevant concentrations.

**Acknowledgements.** The authors thank the Smita Lab for access to the Zeta Sizer for zeta-potential and DLS measurements, the IIT Delhi Central Research Facility (CRF) for SEM imaging, and IIT Delhi High-Performance Computing (HPC) facility for providing computational resources used in this work.

**Author Contribution** Meenakshi, Shalini Gupta, and Gaurav Goel proposed and designed this project together. Meenakshi executed all the experiments, interpreted the results, and prepared the manuscript, figures, tables, and Supplementary Information text. Gaurav Goel reviewed and edited the manuscript.

## Supplementary Information (SI)

### S.1 Calculation of working frequency range

Chassagne et al.[56] provided expressions for the frequency at which mobile charges fully build up a double layer at blocking electrodes ($\omega_p$) and the frequency above which the effects of electrode polarization can be disregarded ($\omega_b$). These expressions are presented in equations S1 and S2:

$$\omega_p = 2\pi f_p = \frac{2K_a D_0}{\Phi_a} \tag{S1}$$

$$\omega_b = 2\pi f_b = \sqrt{\omega_0 \omega_p};\ \ \omega_0 = D_0 K^2 \tag{S2}$$

Here, $D_0$ represents the diffusivity of the ion, $K^{-1}$ represents the Debye length, a is the particle radius, and d is the electrode spacing.

The following values of various parameters were used to calculate $f_p$ and $f_b$ for a suspension of PS particles in PBS buffer:

| | | | | |
|---|---|---|---|---|
| Parameter values: | *d, m* 1.00E-04 | *a, m* 5.00E-07 | $Ka$ 7.04E+02 | $D_0, \frac{m^2}{s}$ 2.00E-09 |
| Electrode polarization frequencies for a suspension of PS paricle in PBS buffer | | | $\omega_p$ , $\frac{1}{s}$ 5.63E+04 | $\omega_b$ , $\frac{1}{s}$ 1.49E+07 |
| | | | $f_p$ , $\frac{1}{s}$ 8.96E+03 | $f_p$ , $\frac{1}{s}$ 2.38E+06 |

**S.2 Frequency-dependent impedance behavior.** We established a lower bound for the operating frequency to minimize system artifacts caused by electrode polarization at the blocking electrodes. Traditional equations for characteristic frequencies in electrokinetic responses do not apply directly to zwitterionic buffers, so we specifically calculated this threshold for PBS. We found that at 9 kHz, charges fully form a double layer near the blocking electrodes ($\omega_p$), while at 2.4 MHz ($\omega_b$), electrode polarization effects become negligible (see Supporting Information for details). Chassagne et al.[56] suggested that for frequencies between $\omega_p$ and $\omega_b$, a correction term can be applied to accurately calculate the permittivity increment ($\Delta\varepsilon_s'$), although no correction is needed for ($\Delta\varepsilon_s''$). As shown in Figure S1, the relative correction term at 1 MHz is less than 10%, indicating that electrode polarization corrections are unnecessary. his conclusion aligns with the previous findings of Khandelwal et al.[37]

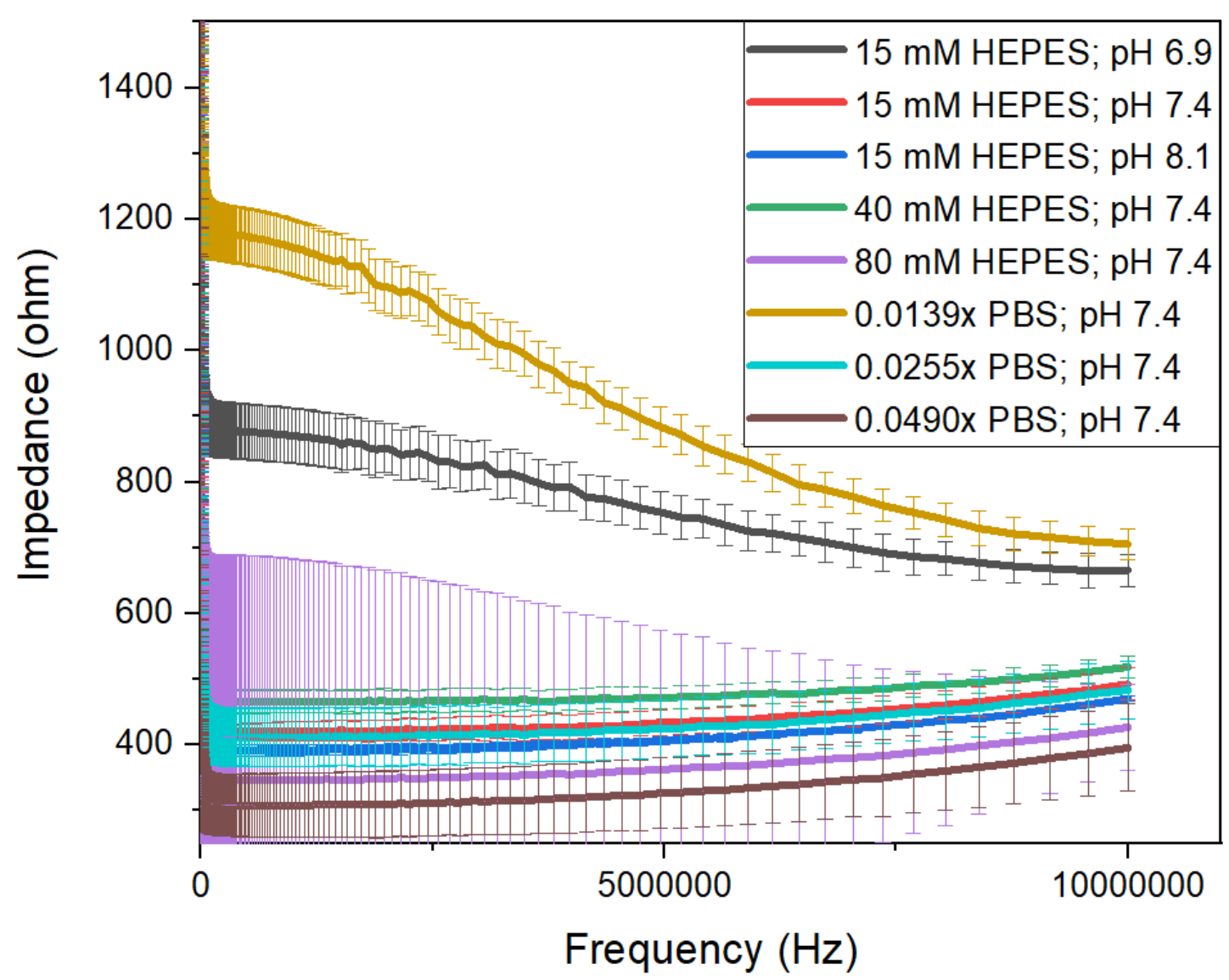

*Figure S1: Impedance spectra as a function of frequency in HEPES and PBS buffer of varying pH and molarity. The results in (a) are reported 5 min after sample loading. Here, the flatness of the HEPES curve (at around 1 MHz) is suggestive of negligible electrode polarization effects.*

Additionally, the frequency-dependent impedance spectra of HEPES and PBS samples showed a plateau (Fig S1) and low phase angle ($\theta < 30^0$) between 10 kHz to 1 MHz, suggesting that the resistive component of system dominates in this frequency range[55]. This allowed us to treat conductance and impedance interchangeably. Based on these considerations, we measured all impedance data at 1 MHz to minimize electrode polarization effects and focused on the bulk properties of the system, including particle-solvent and particle-particle interactions.

**S.3 HEPES speciation and pH dependence**

(a)

(b)

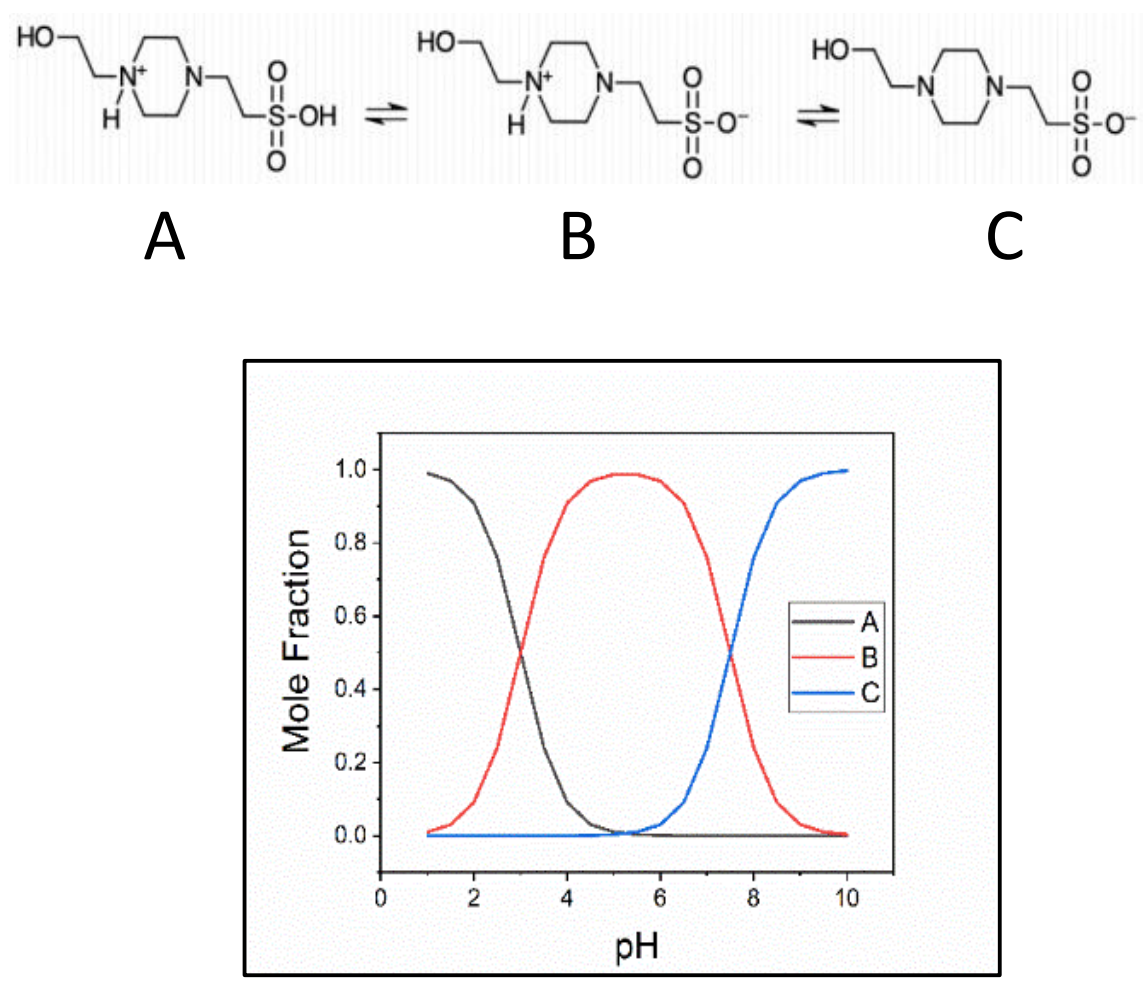

*Figure S2: (a) The different states of HEPES molecule having pKa values at 3 and 7.5, respectively. (b) The graph shows the calculated relative mole fraction of each state as a function of pH.*

**S.4 Oscillatory impedance and buffer comparison**

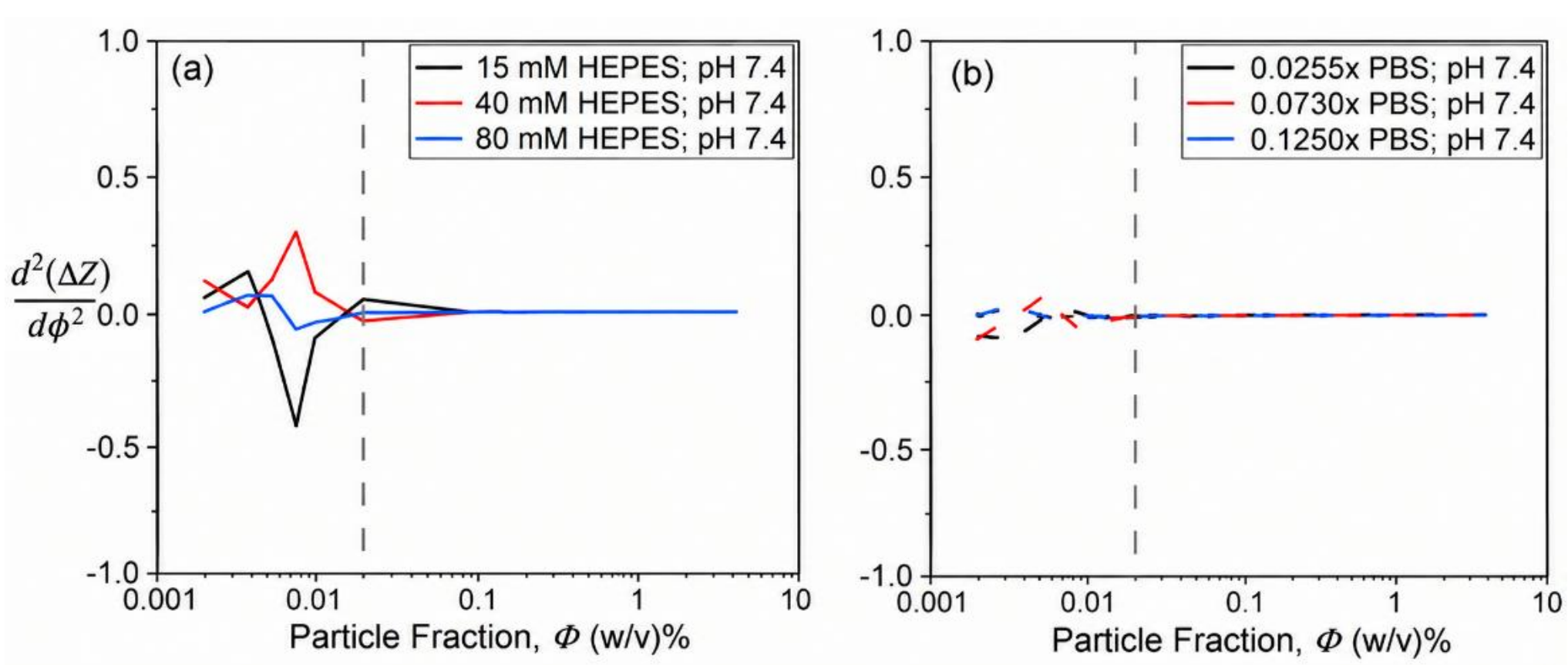

*Figure S3: Second derivative of the impedance change (ΔZ) with respect to particle fraction for (a) 15 mM HEPES (solid black), 40 mM HEPES (solid red), and 80 mM HEPES (solid blue) compared with (b) their conductivity-matched PBS buffers: 0.0255x PBS (dashed black), 0.0730x PBS (dashed red), and 0.1250x PBS (dashed blue) at pH 7.4.*

The comparison highlights differences in oscillatory behavior between HEPES and PBS across different molarities, even under equivalent conductivity conditions.

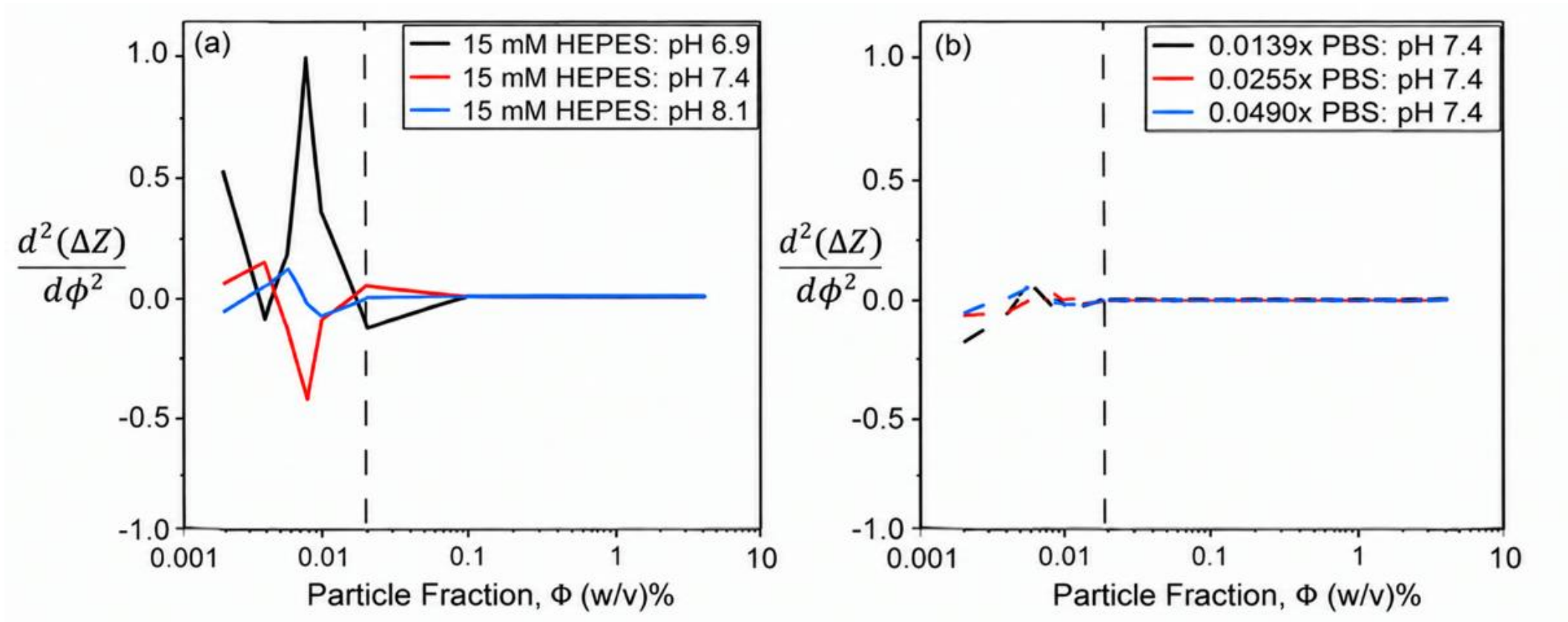

*Figure S4: Second derivative of the impedance change (ΔZ) with respect to particle fraction for (a) 15 mM HEPES at pH 6.9 (solid black line), pH 7.4 (solid red line), and pH 8.1 (solid blue line), compared with (b) their conductivity-matched PBS buffers: 0.0139x PBS (dashed black line), 0.0225x PBS (dashed red line), and 0.0490x PBS (dashed blue line).*

The comparison emphasizes how HEPES at different pH levels, particularly pH 6.9 (where HEPES is in its zwitterionic form), produces stronger oscillatory behavior than the corresponding PBS buffers, even when the conductivity is matched.

## S.5 Atomistic Molecular Dynamics Simulation Methodology

### S.5.1 System Construction

Two atomistic molecular dynamic (MD) systems were constructed to investigate the influence of zwitterionic HEPES species on the electrical double layer (EDL) at a negatively charged polystyrene interface. The simulation suite comprised 15 mM HEPES buffer at pH 6.9 and pH 8.1, enabling direct comparison of the zwitterion-dominated and anion-dominated interfacial regimes under otherwise identical bulk conditions. The polystyrene surface (PS) substrate was modelled as a $10 \times 10 \times 4\ nm^3$ slab using PACKMOL (v20.15.0), comprising 1,000 atactic chains of 150 monomer units each, equilibrated to a bulk density of 0.92 $g/cm^3$. To represent a localized section of a carboxylated microsphere, the slab was asymmetrically functionalized: the interface at z = 4 nm bore 20 carboxylates (COO⁻) groups, yielding a surface charge density $\sigma = -0.20\, e \cdot nm^{-2}$ ($3.2 \times 10^{-2}\, C \cdot m^{-2}$), consistent with experimental values reported for carboxylated latex particles [41]; the opposing face remained uncharged. The carboxylate groups were not covalently tethered to the polymer backbone; instead, they retained lateral mobility to avoid fixed-lattice artefacts and permit spontaneous reorganization in response to the local electrochemical environment.

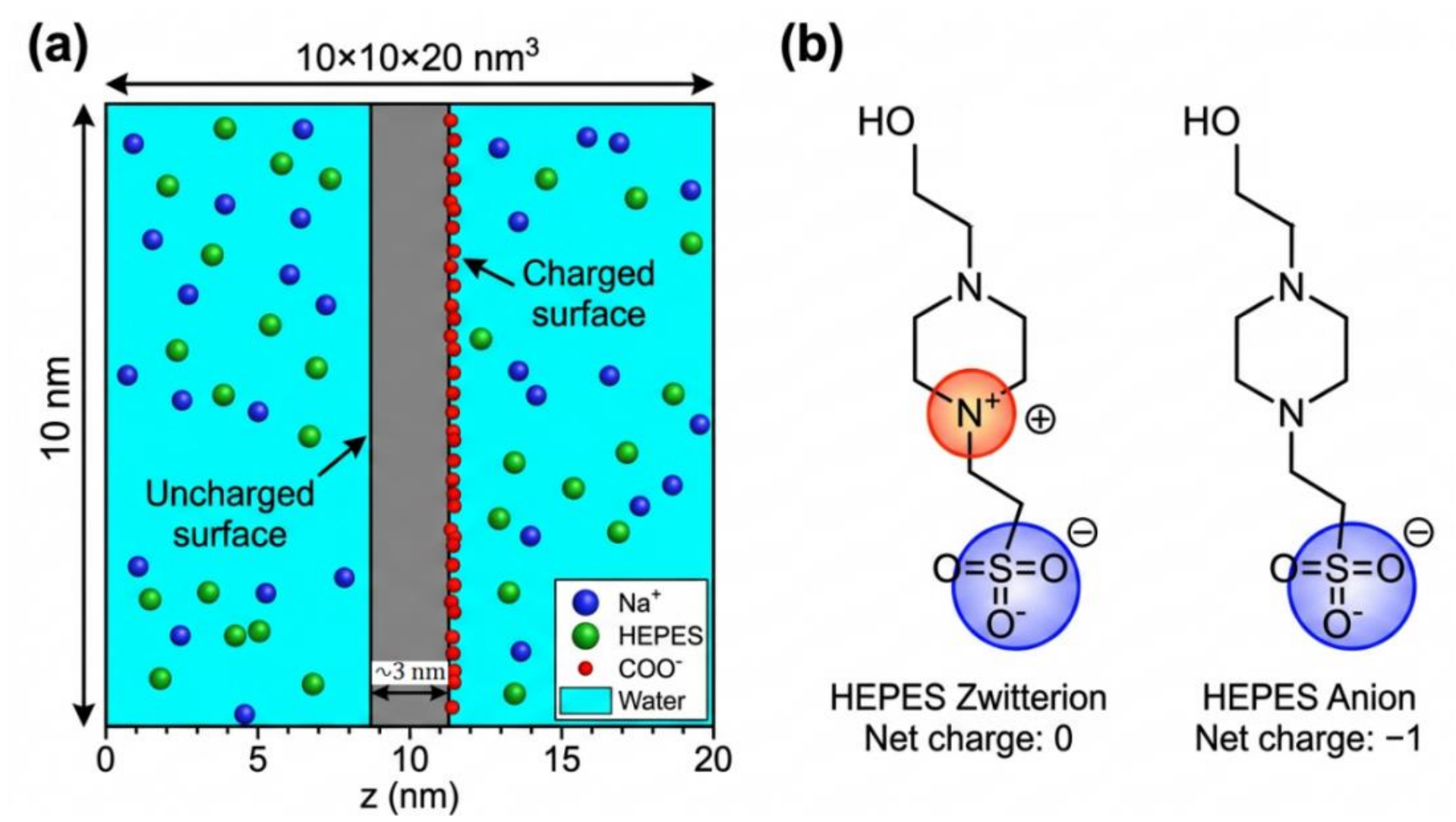

*Figure S5. Simulation setup for the polystyrene–electrolyte interface. (a) Schematic view of the 10 × 10 × 20 nm³ simulation box along the z-direction. The PS slab (grey, ~3 nm) is centred in the box, with the left face uncharged and the right face bearing terminal COO⁻ groups (red). The surrounding aqueous phase contains HEPES molecules (green), $Na^+$ counterions (blue), and water (cyan). (b) Molecular structures of the two HEPES protonation states. The zwitterion (net charge: 0) carries a protonated piperazine nitrogen ($N^+$, orange) and a deprotonated sulfonate group ($SO_3^-$, blue) as two spatially separated charge centres. The anion (net charge: −1) bears a single negative charge localised on the sulfonate moiety ($SO_3^-$, blue).*

The aqueous phase spanned 16.0 nm from the PS surface and contained approximately 119,500 explicit TIP3P water molecules. Protonation states of HEPES were assigned using the Henderson–Hasselbalch equation ($pKa_1 = 3.0, pKa_2 = 7.5$); the resulting pH-dependent speciation is provided in Figure S2. At pH 6.9 (below $pKa_2$), HEPES is predominantly neutral-zwitterionic (~80%), with the anionic form comprising the remainder; at pH 8.1 (above $pKa_2$), the anionic form predominates (~80%), with the zwitterionic fraction reduced to ~20%. System electroneutrality was maintained by the addition of $Na^+$ counterions to balance the surface charge. The final periodic simulation box dimensions were 10.0 × 10.0 × 20.0 $nm^3$. Detailed system compositions and representative simulation snapshots are provided in Table S1 and Figure S1, respectively.

**S.5.2 Simulation Details**

All-atom MD simulations were performed using GROMACS 2020.3 [109, 110]. Molecular interactions were described by the CHARMM36 force field [109], selected for its validated performance in modelling charged interfaces, aqueous electrolytes, and complex biomolecular environments. Water was represented explicitly using the TIP3P potential; HEPES and ionic species were parameterized with compatible CHARMM36 parameters to ensure thermodynamic consistency across all systems.

Each system was energy-minimized using the steepest descent algorithm to a maximum force tolerance of $1 \times 10^{-6}\, kJ \cdot mol^{-1} \cdot nm^{-1}$, followed by a two-stage equilibration protocol: 1.5 ns in the canonical (NVT) ensemble, succeeded by 1.5 ns in the isothermal-isobaric (NPT) ensemble to establish target densities. Temperature was maintained at 298 K using the Nosé–Hoover thermostat ($\tau_{\mathrm{t}} = 1.0\, ps$); pressure was regulated at 1.0 bar using the Parrinello–Rahman barostat ($\tau_{\mathrm{p}} = 2.0\, ps$). System stability was confirmed by monitoring convergence of total energy, box volume, and density throughout the equilibration phase.

Production trajectories of 85 ns were generated in the NPT ensemble using a 2 fs integration time step. Long-range electrostatic interactions were computed via the Particle Mesh Ewald (PME) method [111, 112] with a 1.0 nm real-space cutoff, fourth-order interpolation, and a 0.12 nm Fourier grid spacing. Van der Waals interactions were treated using a Lennard-Jones potential truncated at 1.0 nm, with analytical tail corrections applied to both energy and pressure. All bonds involving hydrogen atoms were constrained using the LINCS algorithm [113]. Periodic boundary conditions were applied in all three dimensions, and atomic coordinates were saved every 10 ps for subsequent structural and dynamical analyses. Molecular visualizations were rendered using VMD [114].

### S.6 Particle morphology and dispersion stability

The particles, visible as bright circular features, are uniformly distributed and retain their spherical shape across all images. No aggregation or interaction between the particles is observed, confirming that at this low concentration, the particles behave independently and remain in a stable, dispersed state. The measurements, highlighted in the bottom-right images, indicate particle diameters ranging between 0.8 $\mu m$ and 1.0 $\mu m$, further supporting their monodisperse nature in the suspension.

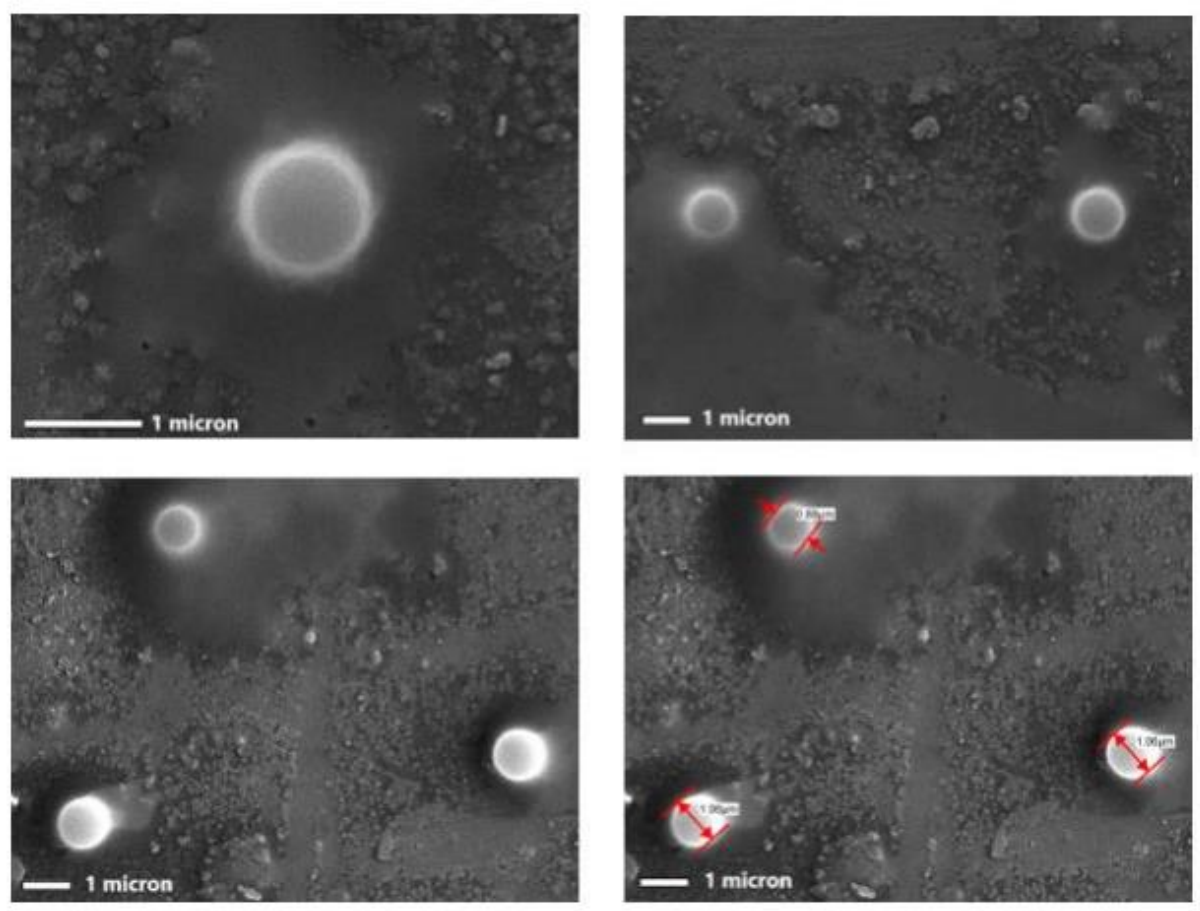

*Figure S6: SEM images depicting carboxylated polystyrene latex particles suspended in a 15 mM HEPES buffer at pH 6.9, with a particle concentration of 0.008 w/v%.*

### S.7 Cluster formation in suspension

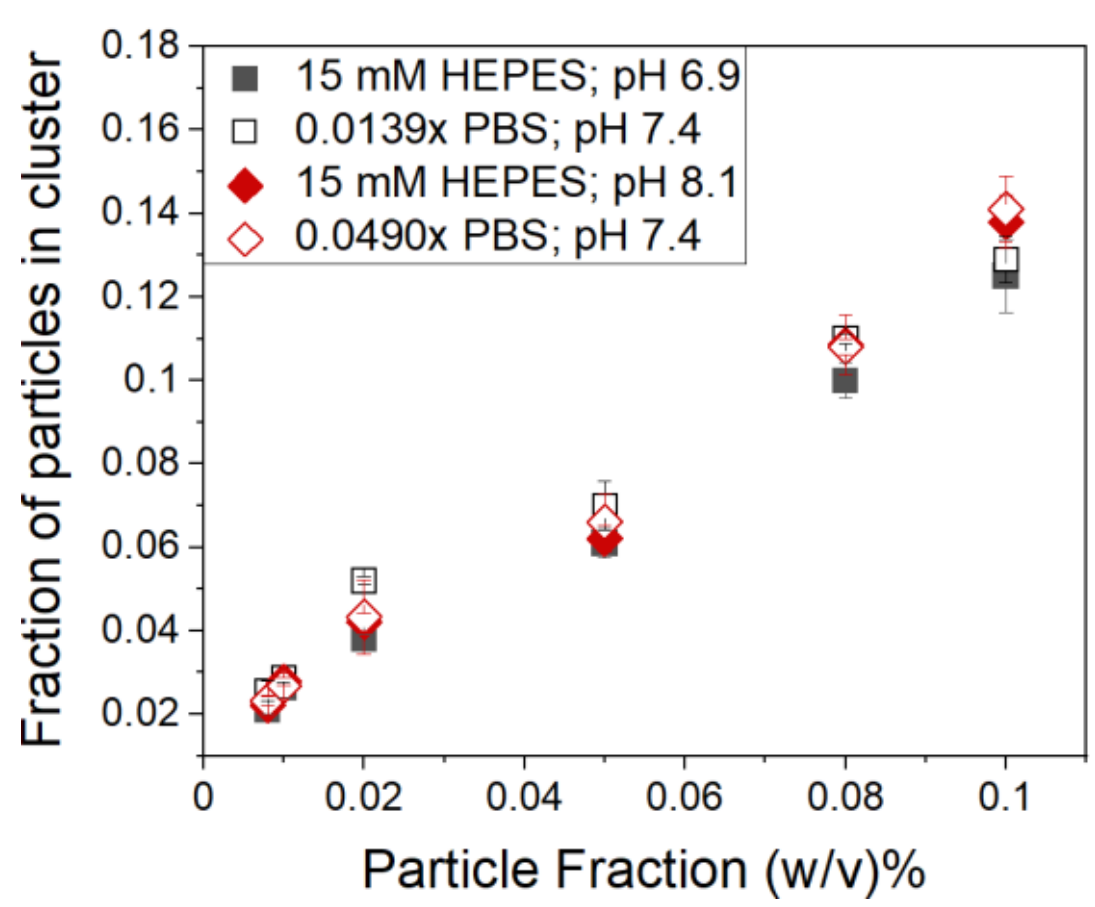

*Figure S7: Fraction of particles in cluster plotted against Φ in suspension for 15 mM HEPES at pH 6.9 (black solid) and 8.1 (red solid), as well as their respective conductivity-matched PBS buffers at 0.0139x (black hollow) and 0.0490x (red hollow). The particle fraction (Φ) in cluster*

*was calculated as the mean from three sets of six images for each sample, with error bars representing the standard deviation.*

**S.8 Comparison of HEPES and PBS microscopy results**

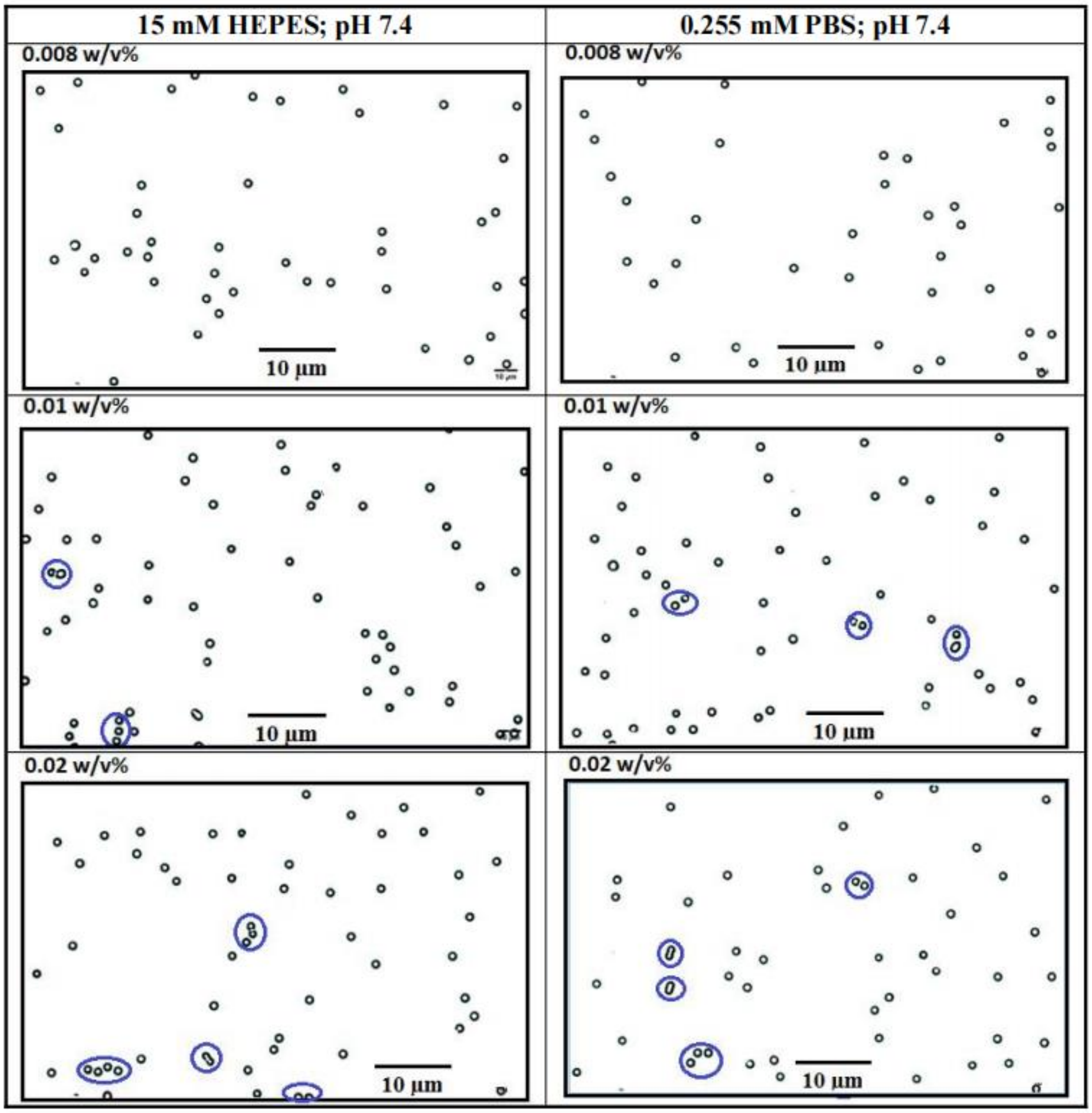

*Figure S8: Microscopy images of 15 mM HEPES at pH 7.4 (left column) and its conductivity-matched 0.0255x PBS at pH 7.4 (right column) for (a), (d) 0.008% w/v, (b), (e) 0.01% w/v and (c), (f) 0.02% w/v particle fractions of PS particles. The blue circles highlight clusters or neighboring particles that are in close proximity, such that no other particles can fit between them.*

**S.9 Zeta potential analysis**

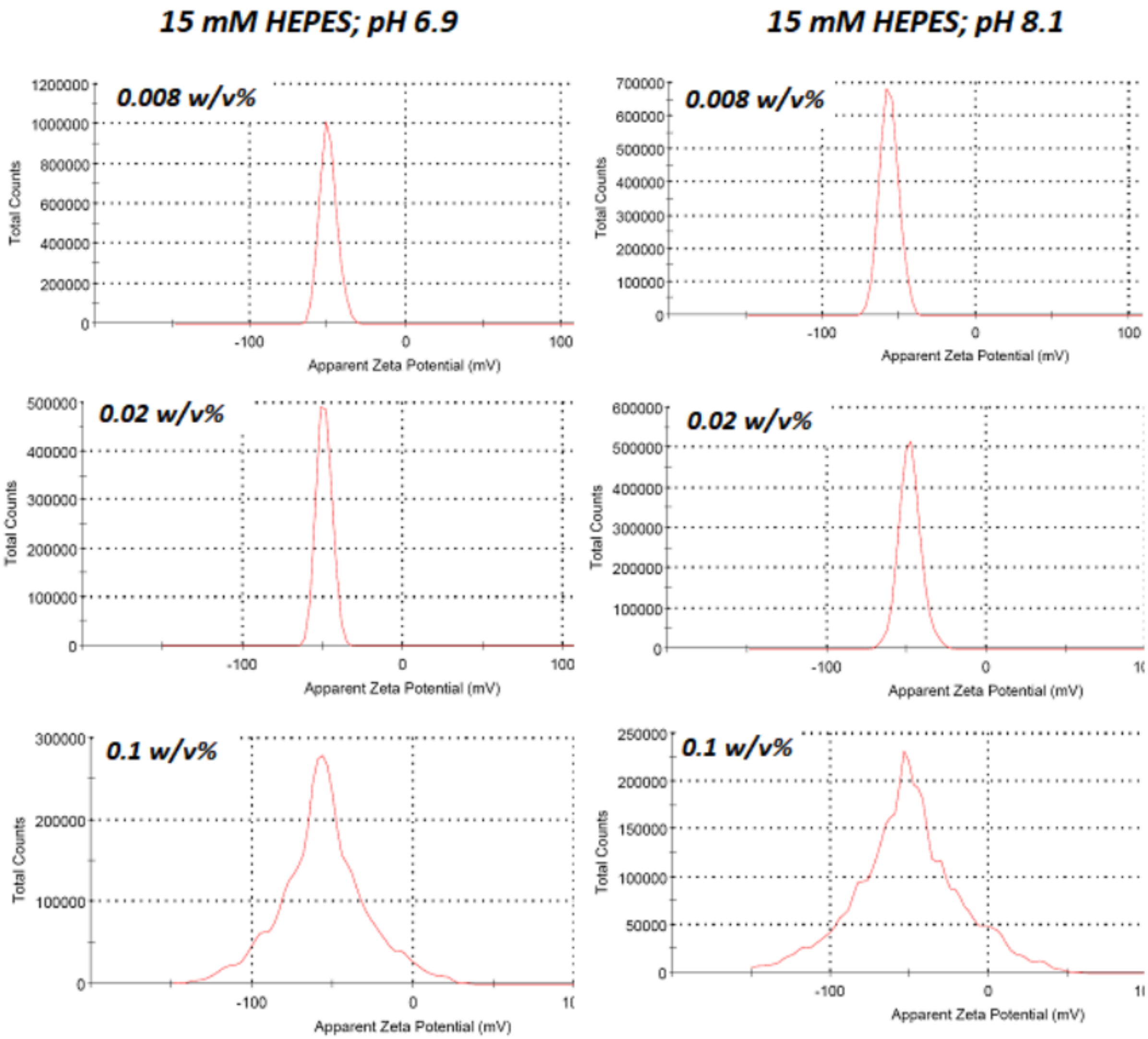

*Figure S9: Zeta potential (mV) values for 15 mM HEPES at pH 6.9 (left column) and pH 8.1 (right column) for three different particle fractions: 0.008 w/v% (Region 1), 0.02 w/v% (critical concentration), and 0.1 w/v% (Region 2). The data show increased polydispersity at higher concentrations, indicating the formation of particle clusters.*

Table S1: Zeta potential and ΔZ values for HEPES at three different buffer concentrations (15 mM, 40 mM, and 80 mM) across pH 6.9 to 8.1, measured at 0.008 w/v% particle fraction.

| Buffer | Particle Fraction (w/v)% | Suspension Conductivity (S/m) | Zeta Potential | | Impedance Measured |
|---|---|---|---|---|---|
| | | | Mean | Std. Dev. | |
| 15 mM HEPES; pH 6.9 | 0.008 | 0.037 | **-63.38** | 0.77 | -115.2 |
| 40 mM HEPES; pH 6.9 | | 0.075 | **-61.8** | 1.15 | -19.86 |
| 80 mM HEPES; pH 6.9 | | 0.115 | **-56.5** | 2.38 | -0.63 |
| 15 mM HEPES; pH 8.1 | | 0.095 | **-54.4** | 2.02 | -4.23 |

The result shows that there is not much change in zeta potential of carboxylated PS particle (-65 mV) at pH 6.9 (where 80% HEPES exists in zwitterionic form) even at higher molarities indicating HEPES doesn't screen but layers at charged surface whereas at pH 8.1, zeta potential value decreases with increase in molarity showing at this pH, HEPES behaves as any other anion.

## S.10 O'Brien's analytical solution

In the limit of a thin double layer, O'Brien et al.[45] provided a measure of the dipole moment generated by the polarization of both the particle core and its double layer as follows:

$$C_o = \frac{\left(2\lambda - \frac{i\omega \varepsilon_{int}}{K_e}\right) - \left(1 - \frac{i\omega \varepsilon_0}{K_e}\right)}{2\left(1 - \frac{i\omega \varepsilon_0}{K_e}\right) + \left(2\lambda - \frac{i\omega \varepsilon_{int}}{K_e}\right)} \quad \text{(S5)}$$

where $C_o(\omega)$ is the dipole coefficient, $\varepsilon_{int}$ is the relative permittivity of the particle core. The complex permittivity $\Delta K_s^*(\omega)$, ignoring $0(\Phi^2)$ and higher order terms, is given by:

$$\frac{\Delta K_s^*(\omega)}{\Phi} = \frac{3}{a^3}\left(K_e - i\omega \varepsilon_0\right) C_o(\omega) \quad \text{(S6)}$$

To compare this theoretical prediction with experimental data, we subtract a frequency-independent offset from only the real part of $\Delta K_s^*(\omega)$, as done by Beltramo et al.[115]. The expression for the double layer conductivity parameter λ was taken from equation 17 of Mangelsdorf et al.[3] when the relevant condiction $\left(exp\left(-\frac{ez_i\xi}{k_BT}\right)>>1\right)$ is satisfied, and from equation 21 (reproduced below) when it is not. Assuming ions with equivalent mobility (or diffusivity), the expression for the double layer conductivity parameter λ is given by

$$\lambda=\frac{2}{Ka}\left(1+\frac{3m}{z^2}\right)\left[\cosh\left(\frac{ez\xi}{2k_BT}\right)-1\right] \quad \text{(S7)}$$

$$m=\frac{z_i\lambda_{avg}\mu}{e} \quad \text{(S8)}$$

$$\mu=\frac{\varepsilon_0 k_BT}{6\pi n_0} \quad \text{(S9)}$$

The average drag coefficient $\lambda_{avg}$ is calculated using a composition-weighted average of individual drag coefficients, $\lambda_i$, given by:

$$\lambda_i=\frac{k_BT}{D_i} \quad \text{(S10)}$$

where $D_i$ is diffusivity of ion.

| Ion | Ionic Diffusivity, $m^2/s$ |
|---|---|
| $Na^+$ | 1.33 x $10^{-9}$ |
| $Cl^-$ | 2.03 x $10^{-9}$ |
| HEPES ion | 5.00 x $10^{-10}$ |

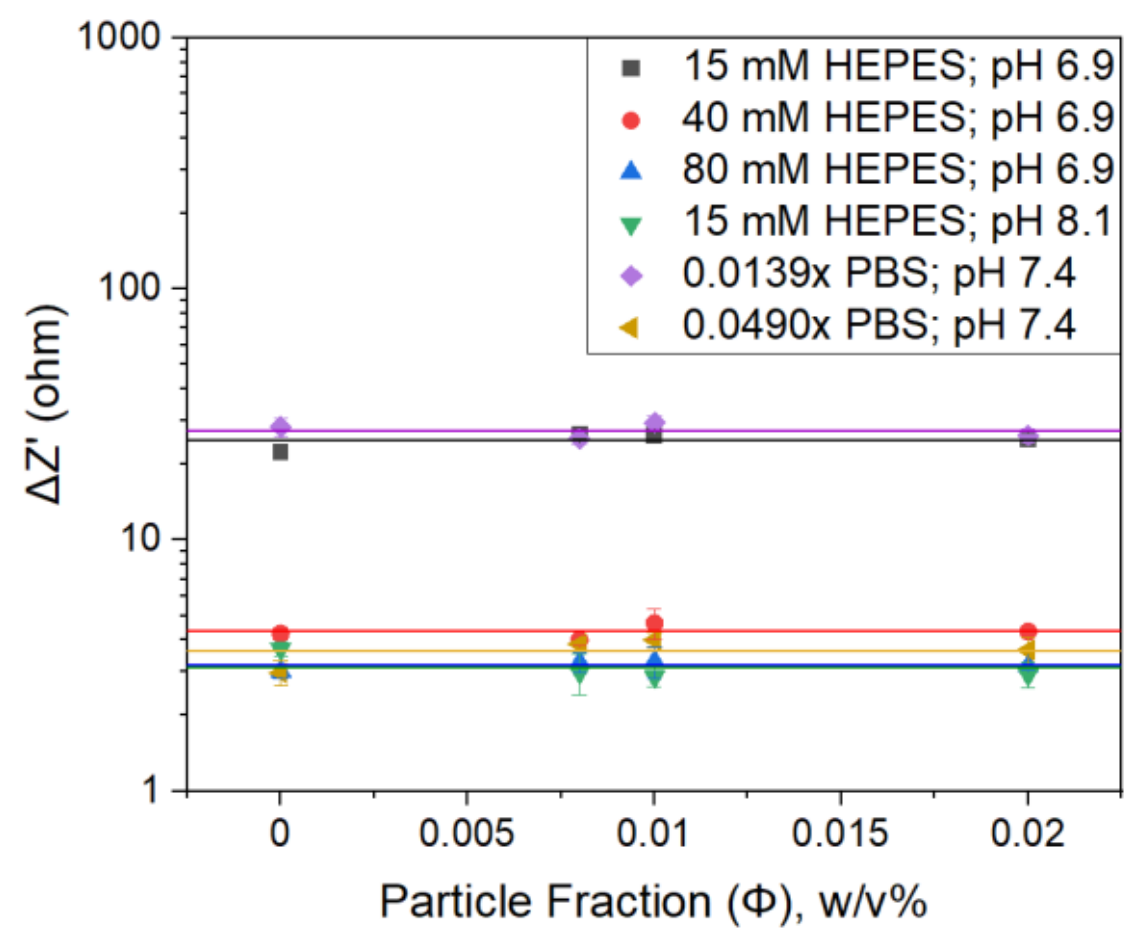

*Figure S10: ΔZ' (offset-corrected change in the real part of impedance) values used to generate offset corrected plots of the impedance change (Figure 3.8) are independent of volume fraction for a given ionic strength. Representative error bars and the average value are also shown. ΔZ' values used to generate offset corrected plots of the conductivity increment (Figure 3.8) are independent of volume fraction for a given ionic strength. Value sensitive error bars and the average value are also shown.*

The offset value is calculated at 100 kHz for all spectra, with the results presented in Figure S9. The corrected data align quantitatively with electrokinetic theory across the entire frequency range, although some discrepancies are noted in the high-frequency shoulder of the conductivity increment. Specifically, above 1 MHz, the corrected conductivity increment decreases as the volume fraction increases for the 1 μm suspensions.

Two key observations can be drawn from the experimental values of $\Delta\sigma'$: (1) the offset is independent of the volume fraction of particles in suspension, and (2) the offset decreases with increasing ionic strength of the suspension, which corresponds to a reduction in double layer thickness. These findings are consistent with previously outlined theories.

(For details: Refer Chassagne et. al.[56])

**S.11 Experimental-theoretical comparison**

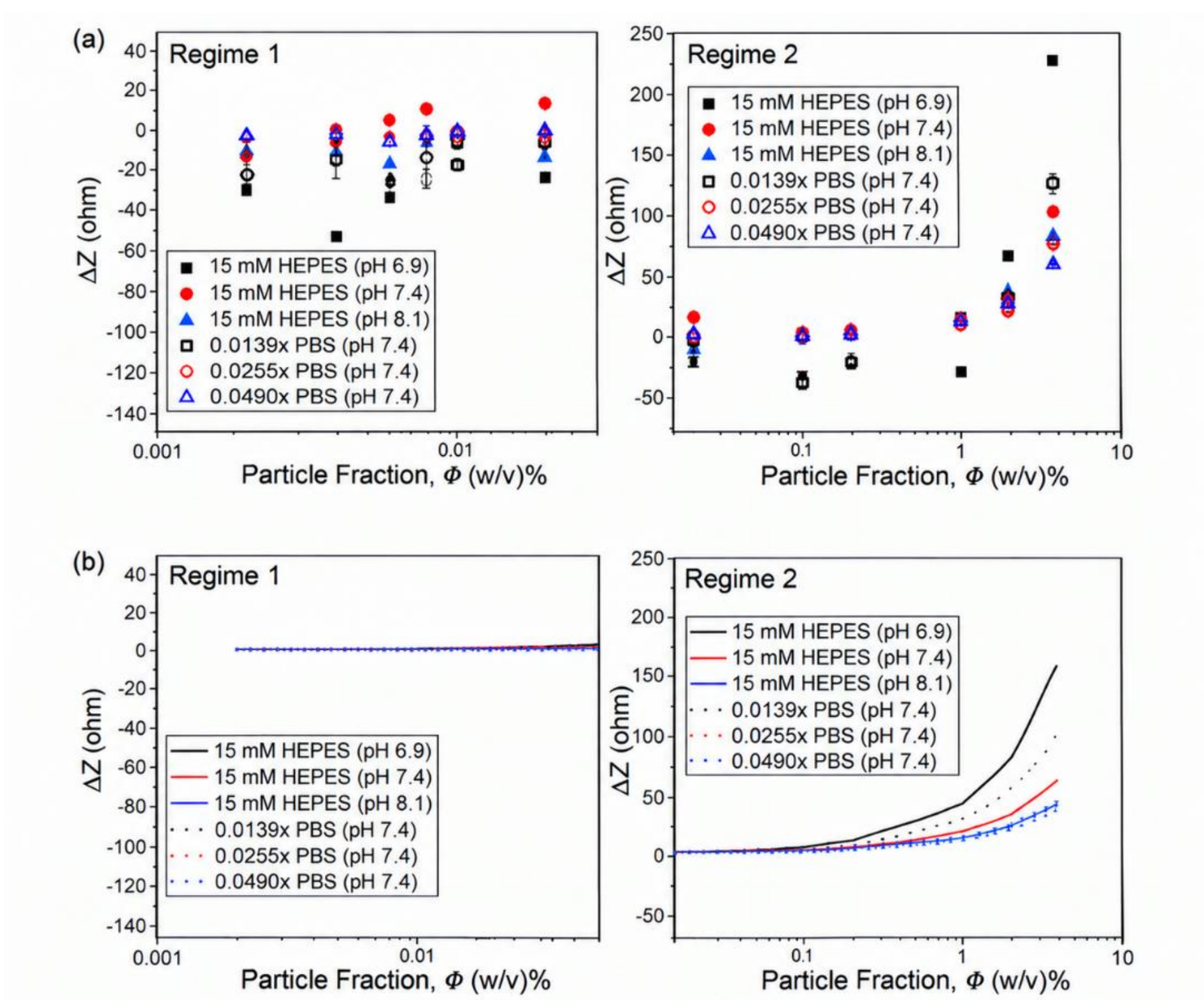

*Figure S11: (a) Plot of the measured ΔZ response for latex suspensions in 15 mM HEPES buffer (solid lines) at pH 6.9 (black), pH 7.4 (red), and pH 8.1 (blue), compared with conductivity-*

*matched PBS buffer (hollow markers) in region 1 (left column) and Region 2 (right column). (b) The ΔZ response contrasted with results from the O'Brien analytical model, shown for both HEPES (solid lines) and PBS (dotted lines), in each region. The comparison highlights the differences between experimental results and theoretical predictions for both buffers across the two regions.*

**S.12 Time-resolved cluster size distribution**

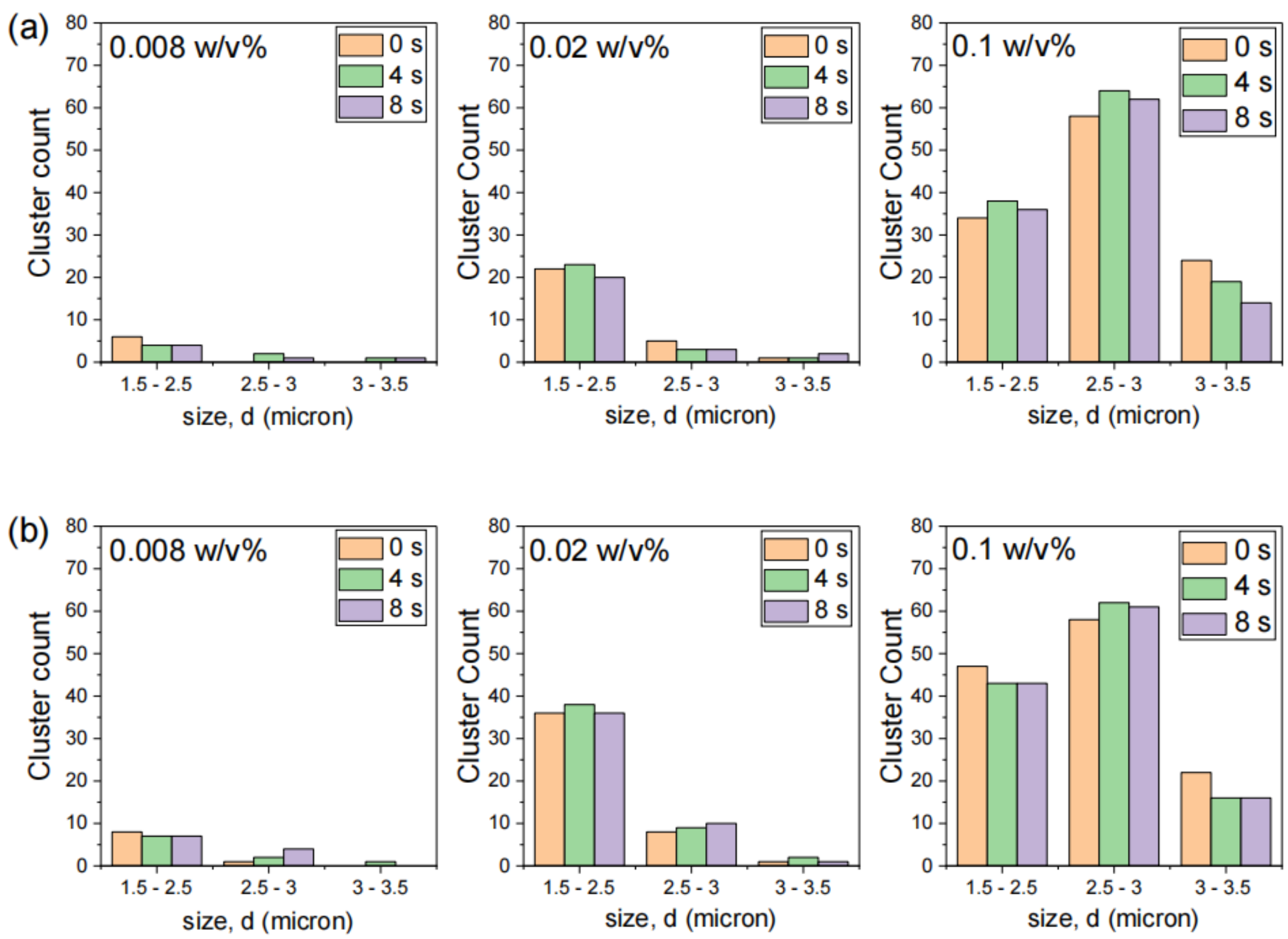

*Figure S12: Size distribution of latex particle clusters at 0 s, 4 s, and 8 s in 15 mM HEPES buffer at pH 7.4 (first row) and its corresponding conductivity matched 0.0225x phosphate buffer (second row), shown by microscopy image analysis at 0.008 w/v% (left column), 0.02 w/v% (middle column), and 0.1 w/v % (right column). To count the clusters, six frames were taken for each particle fraction at different times (after a settling time of 20 minutes)*